\documentclass[lettersize,journal]{IEEEtran}
\def \sn {\textsc{Dehydrator}}

\usepackage{algorithm}
\usepackage{algorithmic}
\usepackage{amsfonts}
\usepackage{amsmath}
\usepackage{textcomp}
\usepackage{array}
\usepackage{xcolor}
\usepackage{epsfig}
\PassOptionsToPackage{hyphens}{url}
\usepackage{svg}
\usepackage{caption}
\usepackage{color}

\usepackage{dingbat}
\usepackage{stfloats}
\usepackage{tabularx}
\usepackage{mdwmath}
\usepackage{multirow}
\usepackage{enumerate}

\usepackage{amssymb}
\usepackage{breakurl}
\usepackage{graphicx}
\usepackage{wasysym}
\usepackage{epstopdf}
\usepackage{subfigure}
\usepackage{mathrsfs}

\usepackage{multirow}
\usepackage{CJKutf8}

\usepackage{url}

\usepackage{threeparttable}
\def\BibTeX{{\rm B\kern-.05em{\sc i\kern-.025em b}\kern-.08em
    T\kern-.1667em\lower.7ex\hbox{E}\kern-.125emX}}
\usepackage{balance}

\begin{document}
\title{\Large \bf \sn{}: Enhancing Provenance Graph Storage via Hierarchical Encoding and Sequence Generation}

\author{Jie Ying,
        Tiantian Zhu*,
        Mingqi Lv,
        Tieming Chen

\IEEEcompsocitemizethanks{
\IEEEcompsocthanksitem This work is supported by the following grants: National Natural Science Foundation of China under Grant No. U22B2028 and 62372410. The Fundamental Research Funds for the Provincial Universities of Zhejiang under Grant No. RF-A2023009. Zhejiang Provincial Natural Science Foundation of China under Grant No. LZ23F020011. 
\IEEEcompsocthanksitem J. Ying, T. Zhu*, M. Lv, and T. Chen are with the College of Computer Science and Technology, Zhejiang University of Technology, Hangzhou 310023, China. E-mail: jieying@zjut.edu.cn, ttzhu@zjut.edu.cn*, mingqilv@zjut.edu.cn, tmchen@zjut.edu.cn. *corresponding author
}
}

\markboth{IEEE Transactions on Information Forensics and Security, ~Vol.~xx, No.~yy, Month~2024}%
{Ying \MakeLowercase{\textit{et al.}}: Dehydrator: Enhancing Provenance Graph Storage via Hierarchical Encoding and Sequence Generation}

\maketitle

\begin{abstract}
As the scope and impact of cyber threats have expanded, analysts utilize audit logs to hunt threats and investigate attacks. The provenance graphs constructed from kernel logs are increasingly considered as an ideal data source due to their powerful semantic expression and attack historic correlation ability. However, storing provenance graphs with traditional databases faces the challenge of high storage overhead, given the high frequency of kernel events and the persistence of attacks. To address this, we propose \sn{}, an efficient provenance graph storage system. For the logs generated by auditing frameworks, \sn{} uses field mapping encoding to filter field-level redundancy, hierarchical encoding to filter structure-level redundancy, and finally learns a deep neural network to support batch querying. We have conducted evaluations on seven datasets totaling over one billion log entries. Experimental results show that \sn{} reduces the storage space by 84.55\%. \sn{} is 7.36 $\times$ more efficient than PostgreSQL, 7.16 $\times$ than Neo4j, and 16.17 $\times$ than Leonard (the work most closely related to \sn{}, published at Usenix Security'23).
\end{abstract}

\section{Introduction}
\label{sec:intro}
With the rapid increase in the scale and complexity of cyberspace over the past decade, cybercrime has grown correspondingly, much like a shadow. 
As of May 2024, there have been 9,478 disclosed data breaches~\cite{itgovernance}. 
Among these, the massive mother of all breaches (MOAB) in January 2024 stands out, with a staggering 26 billion records leaked, including information from platforms such as LinkedIn, Twitter, Weibo, and Tencent~\cite{cybernews_MOAB}. 
Sophisticated and well-funded threat groups seem to demonstrate the ability to infiltrate networks seemingly at will~\cite{india-linked, poland_russian, hacker_ransomware, mitre_groups}, leading to a focus on audit logs to resist post-penetration~\cite{inam2023sok}. 
Auditing is one of the fundamental guarantees of operating system security~\cite{jaeger2022operating}. It is considered an essential condition for detecting vulnerabilities and penetration attempts in any resource-sharing system~\cite{anderson1972computer} and is identified as one of the three pillars of the "Gold Standard" of access control~\cite{Lampson2015SOSP}. 
\par 
More and more evidence suggests that understanding the historical context of attacks through audit logs is crucial for attack hunting and causality analysis~\cite{zhai2006integrating, gu2007bothunter, hassan2020we, chen2021clarion, datta2022alastor}. Provenance graphs constructed from kernel logs are increasingly regarded as an ideal data source for conducting attack investigation due to their powerful semantic expression and attack historic correlation ability~\cite{li2021threat, inam2023sok}. Specifically, a provenance graph is a directed graph structure in which nodes represent system entities (e.g., process, file, and socket) and edges represent system-level events between entities (e.g., write, read, and fork). Provenance tracers continuously capture kernel logs through mature auditing systems to construct corresponding provenance graphs. Then they perform multiple queries on the provenance graph using specific algorithms to hunt potential threats, identify entry points, and determine attacks's impact~\cite{milajerdi2019holmes, milajerdi2019poirot, wang2020you, irshad2021trace, hassan2020tactical, hassan2019nodoze}.
\par
To support these functions, a foundational infrastructure layer responsible for the storage of provenance graphs is essential. This layer must ensure accessibility for upper-layer components. However, four phenomena exist that cause difficulties in storing provenance graphs. (1) Irreversible Growth: To maintain data integrity, provenance tracers only add new data without deleting the existing data~\cite{aldeco2010securing, yagemann2021validating}. This inevitably leads to the continuous expansion of the provenance graph. (2) Rapid Expansion: Due to the complexity of functionalities and the frequency of interactions, auditing frameworks generate a vast amount of kernel logs, with logs exceeding gigabytes per day on a single machine~\cite{ma2018kernel, bates2015trustworthy}. (3) Extended Period: Cyber intrusions targeting government and enterprise systems can persist for extended periods. According to an industry report by TrustWave, the average duration of an intrusion before detection exceeds 188 days~\cite{trustwave}. 
% (4) Need for Unified Analysis: Post initial intrusion, attackers often perform multiple lateral movements~\cite{lateral_movement}. This necessitates the comprehensive analysis of provenance graphs across multiple hosts~\cite{nour2023survey, yagemann2021validating}. 
The three phenomena above result in intolerable storage costs for enterprises and security vendors attempting to centralize the construction of provenance graphs for threat hunting and causality analysis. 
% For example, a partner claims that 100 hosts deployed on Amazon Web Services generate over 85GB of provenance graphs daily. Despite the low storage costs, only the provenance graphs generated within 4 months can be stored simultaneously.
\par
To address the challenge above, many studies start from an intuitive question: \textit{how to store the same provenance graph in a smaller space?} We refer to all efforts aimed at solving this problem as \textbf{Efficient Storage Systems for Provenance Graphs (ESSPGs)}. Methodologically, we categorize existing ESSPGs into two types: Pruning-based and Encoding-based. 
\par
\textbf{Pruning-based ESSPGs}: This type of approach involves pruning parts of the provenance graph based on heuristic rules. LogGC~\cite{lee2013loggc} identifies and deletes events related to temporary and dead nodes, as they do not affect the system state and are useless for causality analysis. CPR~\cite{xu2016cpr} eliminates redundant events between source and target nodes by testing \textit{interleaved flows}, i.e., whether any new inputs have been received at the source between the two system calls. NodeMerge~\cite{tang2018nodemerge} observes that globally read-only files are ineffective for causality analysis and removes events related to these nodes. DPR~\cite{hossain2018dpr} retains only the necessary events for correctly traversing each entity's ancestors (S-DPR) or both ancestors and successors (F-DPR), resulting in a reduced provenance graph. \textit{However, these methods are lossy.} Although they claim that the pruning does not affect upper-layer components, they cannot guarantee that every task will yield correct results. For example, ransomware detection heavily relies on the I/O operations of processes on files, but these events may be considered redundant and thus removed. Therefore, \textit{pruning-based ESSPGs can potentially increase false negatives in the upper-layer components (e.g., threat hunting and causality analysis)}.
\par
\textbf{Encoding-based ESSPGs}: This type of approach represents provenance graphs in a compact encoded form. Compression algorithms, such as Gzip~\cite{deutsch1996gzip} and DeepZip~\cite{goyal2018deepzip}, compress the whole graph as a single file. This approach significantly reduces storage space but cannot support the query requirement of upper-layer components. Graph databases~\cite{neo4j, OrientDB, taitan} store entity and event information in node and edge tables to support queries. Still, they do not consider the specific characteristics of provenance graphs, resulting in limited storage space reduction. SEAL~\cite{fei2021seal} merges repeated fields of events to reduce the storage cost of timestamps. SLEUTH~\cite{hossain2017sleuth} combines encoding techniques, indexed table reference methods, and relative incremental representations, achieving an average storage overhead of only 10 bytes per event. However, the decompression time overhead is insufficient to meet the high demand for queries in causality analysis (e.g., returning the reverse provenance graph for a given point of interest). ELISE~\cite{ding2021elise} and Leonard~\cite{ding2023leonard} combine high-frequency field mapping encoding with deep neural networks (DNN) to store log information. However, their auxiliary components (i.e., calibration tables) take up a lot of storage space thus undermining the objective.
\par
To support the upper-layer components, we identify three key dimensions of ESSPGs within the infrastructure layer (provenance graph storage) from a large body of work in this line of research: (1) \textbf{Content-lossless}, which preserves all data in the provenance graph to avoid leading false negatives in upper-layer components; (2) \textbf{Storage-efficient}, which stores the original provenance graph with minimal storage overhead; (3) \textbf{Query-support}, which handles large-scale query demands from upper-layer components. However, \textit{existing methods fail to satisfy these three requirements simultaneously}.
\par
We propose \sn{}, a provenance graph storage system designed to meet the three requirements mentioned above. First, \sn{} extracts feature information (e.g., unique identifiers and entity names) and structural information (e.g., interactions between entities and timestamps) from the kernel logs collected by the audit framework to construct the provenance graph. This information is stored in the form of node tables and edge tables. Next, \sn{} constructs indexes on the redundant values in fields and replaces them, i.e., field mapping encoding. Then, we observe that causality analysis is an iterative process of querying incoming edges of nodes. Therefore, we treat the provenance graph as a layered directed graph and perform hierarchical encoding. Specifically, \sn{} uses a compact encoding form to store all incoming edge information for a node. Finally, we model querying as a sequence generation task, where the upper-layer component provides input information (e.g., node unique identifier) and returns the corresponding information sequence. This type of task is common in the NLP field. Thus, \sn{} uses DNNs to memorize provenance graph information and constructs error-correction tables to counteract model misprediction. The detailed background and key insights into our design of these mechanisms can be found in Section~\ref{sec:back}. 
\par
For the logs generated by the auditing framework, \sn{} uses field mapping encoding to filter field-level redundancy, hierarchical encoding to filter structure-level redundancy, and finally learns DNN to support batch queries. We evaluated \sn{} on seven datasets totaling over one billion logs. Experimental results show that \sn{} reduces the storage space by 84.55\%. The storage efficiency of \sn{} is 7.36 $\times$ higher than PostgreSQL, 7.16 $\times$ higher than Neo4j, and 16.17 $\times$  higher than Leonard. 
\par
In summary, we make the following contributions:
\begin{itemize}
\item{We propose \sn{}, an efficient provenance graph storage system that overcomes the limitations of existing methods and provides effective storage and querying. \sn{} uses field mapping encoding to filter field-level redundancy, hierarchical encoding to filter structure-level redundancy and finally learns a DNN to support batch querying.}

\item{We built a prototype \sn{} and evaluated it on seven datasets totaling over one billion logs. Experimental results show that \sn{} reduces the storage space by 84.55\%. \sn{} is 7.36 $\times$ more efficient than PostgreSQL, 7.16 $\times$ than Neo4j, and 16.17 $\times$ than Leonard. The code and data in this study will be open-sourced upon publication.}

\item{
We conducted a comprehensive evaluation of \sn{}, exploring the component impact, applicable scenario, and lower bound performance. Additionally, we defined the Latency-to-Storage Ratio, a metric that balances storage overhead and latency, and observed the impact of model capacity on this metric.
}
\end{itemize}

\section{Background \& Insights}
\label{sec:back}
% In this section, we introduce the provenance graph, forensic analysis, and DNN-based storage systems, and present the key insights we have derived for each.
\subsection{Provenance Graph}
\label{sec:back_provenanceGraph}
A typical endpoint detection and response (EDR) system captures streaming kernel logs by invoking an audit framework and represents the provenance graph with a node table and an edge table~\cite{hossain2017sleuth, cheng2023kairos, king2003backtracking}. Specifically, a provenance graph $G(E,\ V)$ is a heterogeneous directed graph consisting of nodes $V$ representing system entities and edges $E$ representing inter-entity interactions. The fields of nodes and edges are carefully selected from raw audit logs, which are lean and critical. 
\par
After observing several threat hunting and attack investigation systems~\cite{cheng2023kairos, zhu2023aptshield, yang2023prographer, milajerdi2019poirot, milajerdi2019holmes, wang2020you, irshad2021trace, hassan2020tactical, hassan2019nodoze}, we found that the fields saved for nodes and edges can be categorized into three types: unique values (e.g., unique identifiers), incremental values (e.g., timestamps), and repetitive values (e.g., node types or event operations). Unique values can be replaced with shorter characters, incremental values with offsets, and repetitive values with indexes. These methods handling \textit{field-level redundancy} are effective in reducing storage space~\cite{fei2021seal, ding2021elise, ding2023leonard, hossain2017sleuth}. However, the storage space for edge information is often significantly larger than the node. For example, when storing the provenance graph for the CADETS group of DARPA TC program~\cite{darpa3_dataset}, we need 42MB of space for the node table but 1,227MB for the edge table ($\sim$ 29.21 $\times$). This is because the operating system handles tasks through numerous system calls. 
% For example, a long-running Firefox process can have hundreds of thousands of related events.
In the provenance graph, this is represented by multiple parallel edges between node pairs, which we refer to as \textit{structure-level redundancy}. 
\subsection{Forensic Analysis}
\label{sec:back_forensicAnalysis}
The goal of forensic analysis using provenance graphs~\cite{milajerdi2019holmes, hassan2019nodoze, hassan2020rapsheet, hassan2020we, xu2022depcomm, fang2022depimpact, zhu2023aptshield, xiong2020conan, king2003backtracking} is to determine the source and scope of the attack, ascertain the extent of disruption, and develop remediation and prevention strategies. 
% By leveraging the rich contextual information provided by provenance graphs, analysts can effectively detect and respond to security events, minimize the impact of attacks, and improve the security of the entire system. 
Typically, analysts perform backtracking from the Point of Interest (POI), iteratively searching for all dependencies related to the current event, similar to a state-constrained reverse breadth-first search (BFS). We also observe that the dependency explosion problem~\cite{hossain2017sleuth, fang2022depimpact, hossain2020combating} causes the number of queries to increase exponentially with the number of iterations. For instance, in the DEPIMPACT dataset~\cite{fang2022depimpact}, obtaining a backtracking graph starting from the POI node (\texttt{/tmp/leaked}) requires over one million events. \textit{We propose modeling the provenance graph as a directed hierarchical graph and performing hierarchical encoding of edges between different hierarchies.} 
% This approach reduces structure-level redundancy caused by parallel edges and supports forensic analysis similar to reverse breadth-first search.

\subsection{DNN-based Storage System}
\label{sec:back_dnn-based_storage_system}
A DNN model is a function capable of learning and representing complex patterns and features within data. Researchers model the storage task as a sequence generation task by building and training a DNN model to capture the data's characteristics and patterns, subsequently using the model to generate the sequence~\cite{goyal2018deepzip, ilkhechi2020deepsqueeze}. During the storage phase, the DNN parameters are optimized on the target data using gradient descent algorithms to fit the data distribution, ultimately retaining only the DNN model while discarding the original data. During the query phase, given an input sequence (i.e., the query statement \texttt{q}), the DNN can iteratively predict the next character and ultimately outputs a string \texttt{s} containing all information. Provenance graph storage is a typical \textit{cold storage} task, where data is queried but not modified.  
Unlike traditional techniques~\cite{stonebraker2018cstore, lamb2012vertica} attempt to balance storage efficiency with processing overhead, we focus instead on minimizing overall data size with query support. 
This emphasis makes time-consuming but storage-efficient DNN-based storage highly applicable. 
% ELISE~\cite{ding2021elise} and LEONARD~\cite{ding2023leonard} have taken initial steps in the field of log storage. 
% However, they have not effectively unified the information-sparse provenance graph with parameter-dense DNNs. The performance of DNNs is too poor (high loss) to achieve efficient storage.
Furthermore, querying DNNs shows better performance than logic-based systems. DNNs inherently support batch queries (e.g., parallelism), which can effectively improve the infrastructure layer performance. 
\par
From Section~\ref{sec:back_provenanceGraph}, we get \textbf{Key Insight \uppercase\expandafter{\romannumeral 1}:} Achieving efficient and lossless storage requires handling both field-level redundancy and structure-level redundancy, with the latter being more critical. From Section~\ref{sec:back_forensicAnalysis} and ~\ref{sec:back_dnn-based_storage_system}, we get \textbf{Key Insight \uppercase\expandafter{\romannumeral 2}:} Hierarchical coding of edges reduces structure-level redundancy, accelerates model training, and supports reverse queries. \textbf{Key Insight \uppercase\expandafter{\romannumeral 3}:} DNN-based storage with specific structures and capacity can better fit encoded provenance graphs, thereby achieving storage-efficient and query-support.

\section{System Design}
\label{sec:sysdesign}
\begin{figure*}[h!t]
\centering
% \vspace{-0.5cm}
\scalebox{0.9}{
\includegraphics[width=\linewidth]{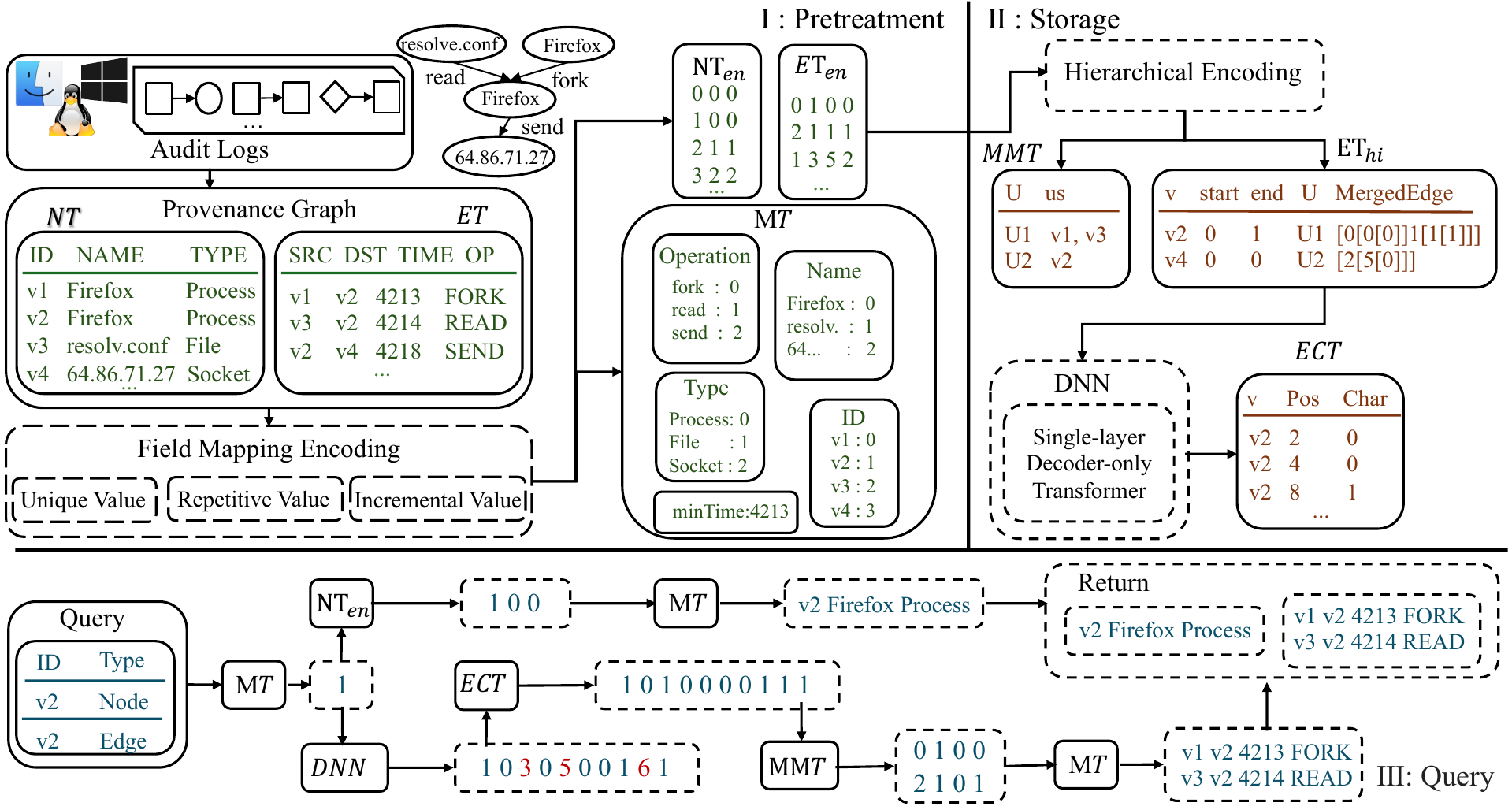}
}
\caption{Overview of \sn{}}
\label{fig:overview}
\vspace{-0.5cm}
\end{figure*}

In this section, we will introduce the design details of each stage of \sn{}. As shown in Figure~\ref{fig:overview}, \sn{} is a three-stage framework (i.e., pretreatment, storage, and query) to achieve efficient storage of provenance graphs and support the upper-layer components.

\par
\subsection{Pretreatment}
\label{sec:sysdesign_preprocessing}
\sn{} leverages mature system auditing frameworks~\cite{etw, linux_subsystem, dtrace_freebsd, redhat} to collect raw logs about system calls from the kernel. Then \sn{}  parses the collected raw logs to build a provenance graph, where nodes represent system entities and edges represent system events. Specifically, \sn{} carefully captures fields using regular expressions. The fields are lean and critical, as shown in Tables~\ref{tab:field_node_edge_table}. \sn{} logically views the unification of two tables (i.e., node table $NT$ and edge table $ET$) as a provenance graph $G(V,\ E)$, which is a common  practice~\cite{cheng2023kairos, yang2023prographer, fang2022depimpact, ding2023leonard}. FOR the node table $NT$, \sn{} preserves IdentiID (the unique identifier for the node), Name, and Type (i.e., file/process/socket). FOR the edge table $ET$, \sn{} preserves SrcID, DstID, TimeStamp (Unix), and Operation (e.g., read/write/execute). Notably, the concatenation of $SrcID$ and $DstID$ is the $IdentiID$ because of the graph structure.
\par
Based on \textbf{Key Insight \uppercase\expandafter{\romannumeral 1}}, \sn{} then addresses three types of field-level redundancy (unique, repetitive, and incremental values) through \textbf{Field Mapping Encoding}. FOR unique and repetitive values, \sn{} replaces them with shorter numerical characters. FOR example, \sn{} uses the numeric character \texttt{0} to replace the frequently occurring string \texttt{Process} in the \textit{Type} field of the node table $NT$. FOR incremental values, \sn{} obtains the minimum value and replaces them with offsets. FOR instance, if a timestamp is \texttt{1522706865} and the minimum value is \texttt{1522706824}, \sn{} will replace \texttt{1522706865} with \texttt{41}. Finally, \sn{} will retain three tables: an encoded node table $NT_{en}$, an encoded edge table $ET_{en}$, and a mapping table $MT$, as shown in Figure~\ref{fig:overview}.
\par
It should be noted that we have retained only the necessary fields to support upper-layer analysis components. However, in practice, \sn{} can modify regular expressions to augment the provenance graph's fields according to specific requirements, such as \textit{PID} or \textit{HostID}. \sn{} can automatically perform field-level redundancy filtering based on the type of newly added fields, demonstrating significant extensibility. In this paper, we exclusively utilize the fields presented in Table~\ref{tab:field_node_edge_table}.

\begin{table}[]
\centering
\caption{Fields in node and edge tables and their corresponding examples and styles.}
\label{tab:field_node_edge_table}
\scalebox{0.95}{
\begin{tabular}{l|c|cc}
\hline
\textbf{Table}        & \textbf{Field} & \textbf{Example}                                                           & \textbf{Style} \\ \hline
\multirow{3}{*}{Node} & IdentiID     & F487A907                                                                   & Unique         \\
                      & Name           & Imapd                                                                      & Repetitive     \\
                      & Type           & File/Process/Socket                                                        & Repetitive     \\ \hline
\multirow{4}{*}{Edge} & SrcID          & A603443D                                                                   & Unique         \\
                      & DstID          & 388D98ED                                                                   & Unique         \\
                      & TimeStamp      & 1522706865                                                                 & Incremental    \\
                      & Operation      & \begin{tabular}[c]{@{}c@{}}Read/Write/Execute\\ Sendto/Recvfr/FORk\end{tabular} & Repetitive     \\ \hline
\end{tabular}
}
\vspace{-0.5cm}
\end{table}

\subsection{Storage}
\label{sec:sysdesign_storing}
In this section, \sn{} achieves efficient storage of the encoded edge table $ET_{en}$ through two steps: (1) Hierarchical Encoding, and (2) Model Training. The former is based on \textbf{Key Insight \uppercase\expandafter{\romannumeral 2}} and filters structure-level redundancy. The latter is based on \textbf{Key Insight \uppercase\expandafter{\romannumeral 3}} and supports batch queries.

\subsubsection{Hierarchical Encoding}
\label{sec:storing_hierarchical_encoding}
\textbf{Elucidation.} Fisrtly, we explain why applying hierarchical encoding to $ET_{en}$ before model training. We model the provenance graph storage task as a sequence generation task, i.e., inputting a sequence and predicting the next character. Besides, we observe that while modern deep neural networks possess complex feature recognition and fitting capabilities, the manifestation of these capabilities requires larger parameter scales and training times. Pre-processed high-dimensional, dense input data can effectively mitigate this issue. This approach has been widely applied in natural language processing~\cite{bengio2000neural, vaswani2017attention}. For instance, when performing text generation tasks, words are first vectorized using word2vec~\cite{mikolov2013word2vec} in the NLP domain. Therefore, we apply hierarchical encoding to increase the information density of input sequences, which significantly improves the training speed and model performance, as shown in Section~\ref{sec:effect_of_components}.
\par
Secondly, we explain why choosing hierarchical encoding. We observe that the direction of most upper-layer analysis components is reversed~\cite{milajerdi2019holmes, hassan2019nodoze, hassan2020rapsheet, hassan2020we, xu2022depcomm, fang2022depimpact, zhu2023aptshield, xiong2020conan, king2003backtracking}. Specifically, when an EDR issues an alert, analysts typically perform multiple queries on the provenance graph based on the infrastructure layer to investigate attack traces and impact scope (i.e., causality analysis), and this process is reversed. Besides, analysts usually use reverse information from the alert point to determine whether the alert is a false positive. We consider this process as a reverse breadth-first search of the provenance graph with conditional constraints. Based on this observation, we model the provenance graph as a directed hierarchical graph and encode different hierarchies separately (i.e., hierarchical encoding), thereby reducing structure-level redundancy.
\par
\begin{figure*}[h!t]
\centering
% \vspace{-0.5cm}
\includegraphics[width=\linewidth]{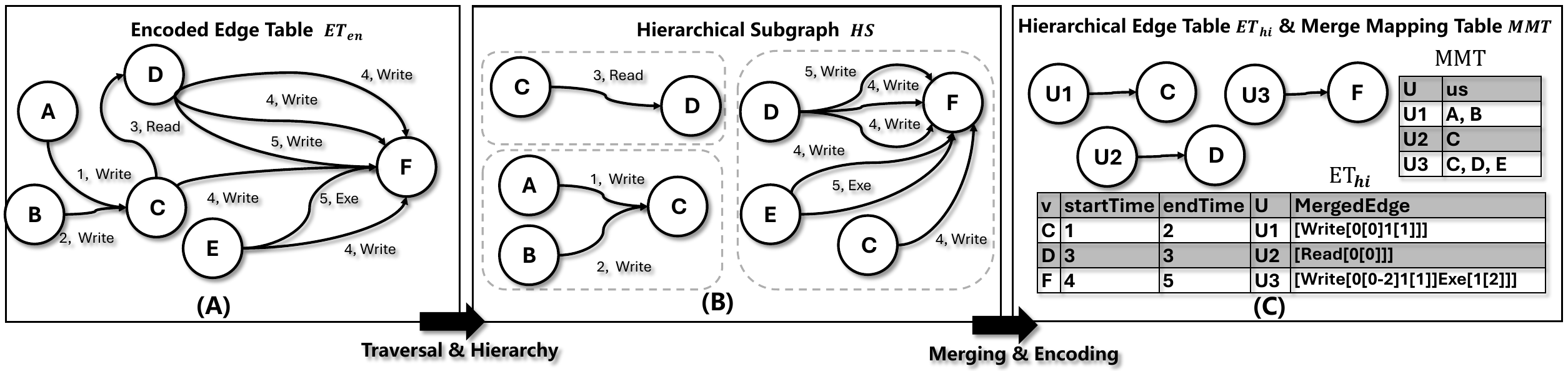}
\caption{Hierarchical Encoding.}
\label{fig:hierarchical_encoding}
\vspace{-0.5cm}
\end{figure*}

\textbf{Procedure.} 
Here, we provide a detailed procedure for hierarchical encoding. In simple terms, \textit{hierarchical encoding is a compact coding form that represents all incoming edges for each node}, an intuitive form of modeling provenance graph queries as reverse breadth-first searches. As shown in Figure~\ref{fig:hierarchical_encoding}, the specific process can be divided into two phases: \textbf{Traversal \& Hierarchy} ($A\to B$), and \textbf{Merging \& Encoding} ($B\to C$).
\par
First, \sn{} traverses the provenance graph, recording all source nodes and incoming edges for each node $v$, as shown in Lines 6 to 8 of Algorithm~\ref{alg:hierarchical_encoding}. \sn{} treats each node $v$ and its incoming edges as a hierarchical subgraph, represented by the gray boxes in part B of Figure~\ref{fig:hierarchical_encoding}. Then, \sn{} save the mapping of all nodes to their source nodes into the $MMT$ and all incoming edges into $MergedEdge$, as shown in Lines 10 to 15 of Algorithm~\ref{alg:hierarchical_encoding}. Next, \sn{} saves the minimum and maximum timestamps from all incoming edges as $startTime$ and $endTime$, as demonstrated in Lines 16 and 17. Finally, \sn{} stores the encoded data into hierarchical edge table $ET_{hi}$ in the format $[v, startTime, endTime, U_v, MergedEdge]$ as shown in Line 18. \sn{} places $v$ at the beginning to serve as a header, supporting the query requirement of DNN-based storage systems. $startTime$ and $endTime$ immediately follow because we observe that temporal constraints are the most common in provenance graph queries by upper-layer analysis components. 
\par
The function $MergeEdges$ combines all edges into a nested list $MergedEdge$, as shown in Part C of Figure~\ref{fig:hierarchical_encoding}. The structure of the nested list is: \textbf{[Operation: [timeOffset: [nodeOffset]]]}, where $timeOffset$=$e.timestamp$-$startTime$, $nodeOffset$=$e.v.Location$ in $U_v$. We designed the list's nesting order based on the amount of redundancy. Firstly, we observed that given a node $v$, the most frequently repeated field among all its incoming edges is \textbf{Operation}. For example, $For\ \forall e \ \in \ E \  where \ \  e.dstID.type=File \to  \ e.Operation \in (Write, Exec)$. The second most repeated field is \textbf{timeOffset}. Due to the complexity and efficiency of modern operating systems, a process node may have hundreds of simultaneously generated incoming edges during initialization. As for \textbf{nodeOffset}, we represent long incremental sequences using shorter format. For example, [1,2,3,5,7,8,9] is represented as [1-3,5,7-9]. 
\par
Finally, \sn{} preserves $MMT$ and $ET_{hi}$, which denotes dense information of all incoming edges $E_v$ of each node $v$ for model training. 
The effectiveness of hierarchical encoding is influenced by the graph structure, a situation we discuss thoroughly in Section~\ref{sec:effect_of_components} and Section~\ref{sec:applicable_scenario}.

% FOR instance, given all incoming edges of file node $v$: $[u1, v, 52, Write]$, $[u2, v, 52, Write]$, $[u3, v, 52, Write]$, $[u4, v, 52, Exec]$, $[u5, v, 55, Write]$, $[u6, v, 57, Exec]$, then $U = [u1, u2, u3, u4, u5, u6], MergedEdge = [Wirte[0[0-2]3[4]], Exec[0[3]5[5]]]$.

\begin{algorithm}[h]  
  \caption{Hierarchical Encoding Algorithm} 
  \label{alg:hierarchical_encoding}  
  \begin{algorithmic}[1]
    \REQUIRE  
        \STATE (1) Encoded Edge Table $ET_{en}$; 
        \STATE (2) Hierarchical Edge Table $ET_{hi}$; 
        \STATE (3) Merge Mapping Table $MMT$;
        \STATE (4) An Empty Dict $Dict$ and two Empty Lists $U_v \& E_v$;
        \STATE (5) Func for Merging Edge Information $MergeEdges$;

    \FOR{$e = (u, v, timestamp, operation) \in ET_{en}$}
     \STATE $Dict[v].append(u)$;
     \STATE $Dict[(u, v)].append(e)$
    \ENDFOR
    \FOR{$v \in Dict.keys()$}
        \FOR{$u \in Dict[v]$}
        \STATE $MMT[U_v].append(u)$
        \STATE $E_v.append(Dict[(u, v)])$
        \ENDFOR
        \STATE $MergedEdge = MergeEdges(E_v)$
        \STATE $startTime = min([e.timestamp \ for \ e \ in \ Dict[(u, v)]])$
        \STATE $endTime = max([e.timestamp \ for \ e \ in \ Dict[(u, v)]])$
        \STATE $ET_{hi}.append([v, startTime, endTime, U_v, $
        \STATE $MergedEdge])$
        \STATE $U_v \gets \emptyset , E_v \gets \emptyset $
    \ENDFOR
    % \STATE \RETURN $ET_{hi} \& MMT$
    \STATE \RETURN $ET_{hi}$ 
  \end{algorithmic}  
\end{algorithm}
% \vspace{-0.5cm}

\subsubsection{Model Training}
\label{sec:model_training}
As described in Section~\ref{sec:back_dnn-based_storage_system}, we model the storage task as a sequence generation task. Specifically, \sn{} functions as a database-like storage system: when a user inputs a query statement \texttt{q}, the DNN outputs a return string \texttt{s} containing all relevant information. Besides, we learn a DNN to support batch queries.
\par
\textbf{Model Selection.} The DNN's output is achieved through iterative input of the generated sequence, i.e., an autoregressive model~\cite{yang2019xlnet, radford2018improving, socher2011parsing} (e.g., LSTM~\cite{hochreiter1997lstm}, GRU~\cite{chung2014gru}, Transformer~\cite{vaswani2017attention}). LSTM and GRU build long-term dependency based on a sequential process, where each hidden state needs to wait for output from previous steps. Transformers build history dependency parallel, which has a better chance of fully utilizing GPU. Standard transformers have deep and heavy model structures whose parameters often reach 300M~\cite{bogoychev2020not, devlin2018bert, tay2020long}. This is because the transformer was originally designed to process linguistic knowledge, requiring the ability to extract high-level semantic features from text. Storage, although modeled as a sequence generation task, focuses more on detecting low-level and repetitive patterns. We believe a single transformer layer is powerful enough to adapt to storage tasks. Furthermore, encoder-only transformers are simpler, have fewer training parameters, and inherently align with sequence generation tasks~\cite{shazeer2019fast, radford2018improving}. \textit{Consequently, we opted for a single-layer, decoder-only transformer.} This opinion is validated in the experiment section.
\par
\textbf{Training \& Error Correction.} 
Initially, \sn{} converts each edge in $ET_{hi}$ into a string $"v, startTime, endTime, U_{v}, MergedEdge"$, where $v$ represents the target node and serves as the query head, $U_{v}$ denotes the index of merged source nodes, and $MergedEdge$ represents the merged edge information of incoming edges. Subsequently, \sn{} directly employs the char2vec method~\cite{cao2016char2vec} to encode the string into numerical vectors. \sn{} then automatically identifies the minimum length $L$ from the query head to the terminator across all strings, and uses this to segment all data. Finally, \sn{} inputs all segmented data into the model for training and terminates the process after reaching the maximum epoch or model fitting.
\par
We model the storage task as a sequence generation task; however, training a zero-error model for a content-rich file is challenging. This conflicts with the need for \textbf{Content-lossless} proposed in Section~\ref{sec:intro}. To address this issue, we employ an Error Correction Table ($ECT$). Specifically, we first test the trained model by sequentially inputting $v$ and comparing the model's final output with the dataset. If inconsistent, we record $(v:[p:c])$ in the $ECT$, where $v$ represents the query head, $p$ indicates the error position and $c$ denotes the correct character. Finally, we preserve this $ECT$ to correct the model's generated content when supporting query requirements of upper-layer analysis components. This mechanism is similar to existing work~\cite{ding2023leonard}.

\subsection{Query}
\label{sec:sysdesign_query}
\sn{}, as an efficient infrastructure layer, supports query requirements of upper-layer analysis components. We describe the process in the context of Figure~\ref{fig:overview}. FOR node information, given a node's $IdentiID$ \textbf{v2}, \sn{} obtains its index value \textbf{1} via $MT$ mapping ($IndexValue=MT(IndentiID)$). Then \sn{} searches in the encoded node table $NT_{en}$ by index, and retrieves node information \textbf{(v2 Firefox Process)} through $MT$ inverse mapping ($Information=MT\_REVERSE(NT_{en}(IndexValue))$). FOR edge information, \sn{} inputs the index value into the model, generating the corresponding sequence \textbf{[1 0 3 0 5 0 0 1 6 1]} through autoregression ($Seq=DNN(IndexValue)$). Subsequently, \sn{} employs $ECT$ for error correction, obtaining a fully accurate sequence \textbf{[1 0 1 0 0 0 0 1 1 1]} ($AccSeq=ECT(Seq)$). Finally, \sn{} performs hierarchical decoding and $MT$ inverse mapping on the generated sequence to retrieve readable information \textbf{[v1 v2 4213 FORK], [v3 v2 4214 READ]} for all incoming edges ($Information=MT\_REVERSE(HierDecode(AccSeq))$).

\section{Evaluation}
\label{sec:evaluation} 
In this section, we first introduce the experimental protocol, including the environment, baselines, metrics, datasets, and hyperparameters. Subsequently, we provide a detailed assessment of the costs with storage and query of provenance graphs. Then, we demonstrate the significance of each component of \sn{}. Furthermore, we determine the applicable scenario and lower bound performance of \sn{}. Finally, we explore the impact of model selection and capacity on \sn{}. 
In summary, we aim to answer the following questions:
\begin{itemize}
\item{\textbf{RQ1:} What is the performance of \sn{} on provenance graph storage?}
\item{\textbf{RQ2:} How efficient is \sn{} in supporting provenance graph queries?} 
\item{\textbf{RQ3:} How important is each component of \sn{}?}
\item{\textbf{RQ4:} What is the applicable scenario and lower bound performance of \sn{}?}
\item{\textbf{RQ5:} What is the impact of model selection and capacity on \sn{}?}
\end{itemize}

\subsection{Experimental Protocol}
\label{sec:eval_experimental_protocol}
\textbf{Environment.} \sn{} is implemented using Python 3.9 with PyTorch 1.13.1 and scikit-learn 1.2.0, used for training DNN models, respectively. Our experiments are all carried out on a server running Ubuntu 20.04 64-bit OS with AMD EPYC 7513 32-Core Processor, 256GB memory, and RTX A6000 GPU.
\par
\textbf{Baselines.}We have selected existing state-of-the-art systems as baselines for comparison with \sn{}: PostgreSQL~\cite{postgresql} (a relational database),  Neo4j~\cite{neo4j} (a graph database), and LEONARD~\cite{ding2023leonard} (a DNN-based storage system). 
\par
\textbf{Metrics.} We define the storage overhead (bytes) of the pre-processed provenance graph as $BP_{pre}$, the post-processed as $BP_{post}$, and the storage latency as $T_{s}$. Specifically, $BP_{pre}$ represents the bytes of the $ET$, while $BP_{post}$ is the bytes for three files ($MMT$, $DNN$, and $ECT$). Considering that $ET$ occupies more than 95\% of the space in the provenance graph, we prefer to evaluate the efficiency of \sn{} in storing $ET$, thus equating $BP_{pre}$ to $ET$. 
% The Storage-Compression Ratio (SCR) is then defined as $SCR = (BP_{pre} - BP_{post}) / BP_{pre}$, and the Latency-to-Storage Ratio (LSR) is defined as $LSR = (BP_{pre} - BP_{post}) / T_{s}$. In addition, we define the latency of querying as $T_{q}$. 
The Latency-to-Storage Ratio (LSR) is defined as $LSR = (BP_{pre} - BP_{post}) / T_{s}$. 
% In addition, we define the latency of querying as $T_{q}$. 

% Since the $ET$ occupies over 95\% of the space in the provenance graph, and the field-level redundancy filtered by \sn{} on the $NT$ is constant, \sn{} focuses more on evaluating the efficiency of $ET$ storage.

\par
\textbf{Datasets.} 
We evaluate \sn{} on a dataset containing over one billion raw logs. As shown in Table~\ref{tab:datasets}, this dataset comprises seven provenance graphs (G1-G7). G1-G4 are sourced from the CADETS (G1), THEIA (G2), and TRACE (G3, G4) groups of DARPA TC E3~\cite{darpa3_dataset}, G5-G6 are from the TRACE group of DARPA TC E4, G7 is a subset of the DEPIMACT dataset~\cite{fang2022depimpact}. G1-G6 were collected using SPADE~\cite{gehani2012spade}, while G7 is  collected using Sysdig~\cite{sysdig}. 
% We collected the G7 and G8 datasets under busy and idle host environments, respectively. 
The second column, "Log Size," indicates the size of the raw log files collected by the capture layer~\cite{gehani2012spade, sysdig}, i.e., JSON files. The third column, "Graph Size," represents the size of the provenance graphs, i.e., $BP_{pre}$. The fourth column, "\# Nodes," and the fifth column, "\# Edges," denote the number of nodes and edges in the provenance graph. \textit{Regarding the average edges of provenance graphs ($\sim$ 17,754,566), we are 9.6 $\times$ larger than Leonard ($\sim$ 1,849,833).} 
\par
\textbf{Hyperparameters.} Unless otherwise specified, \sn{} uses the same hyperparameters. Specifically, \sn{} employs the Single-layer Decoder-only Transformer model described in Section~\ref{sec:model_training}. The default training settings for the dataset are batch size 4,096, Adam optimizer, learning rate 0.001, and maximal training epoch 5, with an early stopping mechanism. Additionally, the length of segments for different provenance graphs in the dataset is adaptive and different. Finally, we design several models with different capacities (i.e., embedding dimension, number of attentional heads, and feedforward dimension) to evaluate the impact of model capacity on \sn{}, as shown in Table~\ref{tab:models_overview}.

\begin{table}[]
\centering
\caption{Overview of Datasets}
\label{tab:datasets}
\scalebox{0.99}{
\begin{tabular}{ccccc}
\hline
\textbf{Label} & \textbf{\begin{tabular}[c]{@{}c@{}}Log Size \\ (GB)\end{tabular}} & \textbf{\begin{tabular}[c]{@{}c@{}}Graph Size \\ (MB)\end{tabular}} & \textbf{\begin{tabular}[c]{@{}c@{}}\#\\ Nodes\end{tabular}} & \textbf{\begin{tabular}[c]{@{}c@{}}\#\\ Edges\end{tabular}} \\ \hline
G1             & 38.34                                                             & 1226.72                                                             & 414,518                                                     & 13,244,643                                                  \\
G2             & 85.19                                                             & 2750.08                                                             & 1,144,475                                                   & 29,655,652                                                  \\
G3             & 25.56                                                             & 712.92                                                              & 812,083                                                     & 7,628,155                                                   \\
G4             & 26.89                                                             & 713.84                                                              & 816,730                                                     & 7,717,378                                                   \\
G5             & 32.05                                                             & 1152.30                                                             & 1,001,005                                                   & 12,536,153                                                  \\
G6             & 33.06                                                             & 1164.28                                                             & 966,490                                                     & 12,691,262                                                  \\
G7             & 31.89                                                             & 3470.72                                                             & 432,379                                                     & 40,808,719                                                  \\ \hline
\textbf{Avg}   & 38.99                                                             & 1598.69                                                             & 798,240                                                     & 17,754,566                                                  \\ \hline
\end{tabular}
}
\vspace{-0.2cm}
\end{table}

\begin{table}[]
\centering
\caption{Models with Different Capacities. Column 1 is the capacity (M is short for MEGABYTE) of the models. Columns 2 through 4 are adjustable hyperparameters. Column 5 is the number of parameters.}
\label{tab:models_overview}
\scalebox{0.97}{
\begin{tabular}{ccccc}
\hline
\textbf{Capacity (M)} & \textbf{\begin{tabular}[c]{@{}c@{}}Embedding\\ Dimension\end{tabular}} & \textbf{\begin{tabular}[c]{@{}c@{}}Attention\\ Head\end{tabular}} & \textbf{\begin{tabular}[c]{@{}c@{}}Feedfoward\\ Dimension\end{tabular}} & \textbf{\begin{tabular}[c]{@{}c@{}}Parameters\\ (K)\end{tabular}} \\ \hline
C1 (0.08)          & 32                                                                     & 1                                                                 & 128                                                                     & 18                                                                \\
C2 (0.28)          & 64                                                                     & 2                                                                 & 256                                                                     & 70                                                                \\
C3 (1.1)           & 128                                                                    & 4                                                                 & 512                                                                     & 271                                                               \\
C4 (2.4)           & 192                                                                    & 6                                                                 & 768                                                                     & 600                                                               \\
C5 (4.1)           & 256                                                                    & 8                                                                 & 1,024                                                                   & 1,070                                                             \\
C6 (9.2)           & 384                                                                    & 12                                                                & 1,536                                                                   & 2,380                                                             \\ \hline
\end{tabular}
}
\vspace{-0.5cm}
\end{table}

\begin{table*}[]
\centering
\caption{Storage Overhead and Latency of Storing Provenance Graphs for Different Systems. $BP_{post}$ and $T_{s}$ denote the disk usage and latency of storing the provenance graph. $MT$, $DNN$, and $ECT$ represent the storage overhead for the merge mapping table, model, and error correction table. HE, Train, and Correct represent the time overhead for hierarchical encoding, model training, and error correction.}
\label{tab:storage_performance}
\scalebox{0.99}{
\begin{tabular}{c|cc|cc|cccc|cccccc}
\hline
\multirow{3}{*}{\textbf{Dataset}} & \multicolumn{2}{c|}{\textbf{PostgreSQL}}                                                                      & \multicolumn{2}{c|}{\textbf{Neo4j}}                                                                        & \multicolumn{4}{c|}{\textbf{Leonard}}                                   & \multicolumn{6}{c}{\textbf{Dehydrator}}                                      \\
                                  & \multirow{2}{*}{\begin{tabular}[c]{@{}c@{}}$BP_{post}$ \\ (MB)\end{tabular}} & \multirow{2}{*}{$T_{s}$ (s)} & \multirow{2}{*}{\begin{tabular}[c]{@{}c@{}}$BP_{post}$\\ (MB)\end{tabular}} & \multirow{2}{*}{$T_{s}$ (s)} & \multicolumn{2}{c}{$BP_{post}$ (MB)} & \multicolumn{2}{c|}{$T_{s}$ (s)} & \multicolumn{3}{c}{$BP_{post}$ (MB)}       & \multicolumn{3}{c}{$T_{s}$ (s)} \\ \cline{6-15} 
                                  &                                                                              &                              &                                                                             &                              & $DNN$  & \multicolumn{1}{c|}{$ECT$}  & Train          & Correct         & $MMT$ & $DNN$ & \multicolumn{1}{c|}{$ECT$} & HE      & Train    & Correct    \\ \hline
G1                                & 1,362                                                                        & 33                           & 1,367                                                                       & 17                           & 0.9    & \multicolumn{1}{c|}{3,592}  & 15,273         & 7,628           & 8.8   & 1.1   & \multicolumn{1}{c|}{51}    & 55      & 660      & 429        \\
G2                                & 3,034                                                                        & 76                           & 2,767                                                                       & 28                           & 0.9    & \multicolumn{1}{c|}{5,896}  & 38,054         & 16,324          & 21.2  & 1.1   & \multicolumn{1}{c|}{135}   & 130     & 1,763     & 1,056       \\
G3                                & 786                                                                          & 20                           & 1,037                                                                       & 15                           & 0.9    & \multicolumn{1}{c|}{2,217}  & 8,913          & 4,517           & 26.8  & 1.1   & \multicolumn{1}{c|}{252}   & 79      & 2,709     & 1,724      \\
G4                                & 790                                                                          & 21                           & 1,039                                                                       & 17                           & 0.9    & \multicolumn{1}{c|}{2,419}  & 11,295         & 5,824           & 30.4  & 0.1   & \multicolumn{1}{c|}{292}   & 91      & 1,269     & 2,070      \\
G5                                & 1,287                                                                        & 32                           & 1,306                                                                       & 19                           & 0.9    & \multicolumn{1}{c|}{3,659}  & 14,078         & 7,058           & 33.1  & 1.1   & \multicolumn{1}{c|}{345}   & 111     & 1,924     & 2,300      \\
G6                                & 1,291                                                                        & 33                           & 1,319                                                                       & 21                           & 0.9    & \multicolumn{1}{c|}{3,176}  & 12,942         & 6,091           & 28.3  & 2.4   & \multicolumn{1}{c|}{359}   & 102     & 1,913     & 1,888      \\
G7                                & 4,177                                                                        & 103                          & 3,556                                                                       & 33                           & 0.9    & \multicolumn{1}{c|}{7,001}  & 43,206         & 20,433          & 9.8   & 1.1   & \multicolumn{1}{c|}{131}   & 166     & 879      & 1,122      \\ \hline
\textbf{Avg}                      & 1,818                                                                        & 45                           & 1,770                                                                       & 21                           & \multicolumn{2}{c|}{3,991}           & \multicolumn{2}{c|}{30,233}       & \multicolumn{3}{c|}{\textbf{247}}          & \multicolumn{3}{c}{3,205}       \\ \hline
\end{tabular}
}
\vspace{-0.1cm}
\end{table*}

\subsection{Storing Provenance Graphs}
\label{sec:rq1}
To address RQ1, we evaluate the storage overhead and latency of \sn{} and different baselines (PostgreSQL, Neo4j, and Leonard) operating on various datasets with metrics $BP_{post}$, $T_{s}$. For PostgreSQL, we use the built-in COPY command to store the provenance graphs from CSV files~\cite{postgresql-copy-command}. For Neo4j, we utilize the \textit{neo4j-admin import} tool to store the provenance graphs~\cite{Neo4j-admin-import}, specializing in storing large graphs. For Leonard\footnote{https://github.com/dhl123/Leonard}, we modify the parsing component to take the $ET$ as input while keeping other components unchanged. We also increase $\#Parameters$ of the multilayer LSTM to $64K$ to align ours. We email the authors and will also open-source this modified version of Leonard upon publication. 
\par
Table~\ref{tab:storage_performance} shows the storage overhead and latency of different techniques for storing provenance graphs. As can be observed, the average storage overhead for the relational database PostgreSQL is 1,818M, while for the graph database Neo4j it is 1,770M, both slightly larger than the original provenance graphs ($\sim$ 1,598M). These traditional databases build indices and maintain metadata to enhance query performance when storing data. Leonard performs the worst, with an average storage overhead of 3,991M, which is 2.5 $\times$ larger than the original provenance graphs. Leonard fails to achieve efficient storage of provenance graphs, strikingly different from the performance reported in the original paper~\cite{ding2023leonard}. \textit{\sn{} demonstrates the best storage performance, at only 247M, which is 7.36 $\times$ smaller than PostgreSQL, 7.16 $\times$ smaller than Neo4j, and 16.17 $\times$ smaller than Leonard.}
\par
\sn{} significantly outperforms Leonard, and there are four reasons for this. (1) Evaluation Metric. Leonard evaluates storage overhead on the gzip-compressed $ECT$ file, where a 3592M $ECT$ is compressed to a 68M \texttt{gz file}. However, the storage system needs to decompress this \texttt{gz file} and load it into memory to support queries, which requires significant time (138s) and memory overhead (3.6G). We therefore use the size of raw $ECT$ (JSON File) as an evaluation metric for storage overhead. (2) Dataset Complexity. The average number of edges and degrees in our dataset ($\sim$ 17,754,566, $\sim$ 22.24) are 9.6 $\times$ and 11.7 $\times$ those of Leonard ($\sim$ 1,849,833, $\sim$ 1.89) respectively. In non-Euclidean space, the complexity of the graph structure may exceed Leonard's processing boundaries. Furthermore, we only involve 4 fields (SrcID, dstID, Operation, and Timestamp), while Leonard has 13. We believe that Leonard's reported efficient storage capability stems mainly from filtering field-level redundancy but ignoring structure-level redundancy. (3) Hierarchical Encoding. We can view hierarchical encoding as an operation that aggregates information to increase data density. This heuristic approach helps the model avoid learning patterns directly from low-dimensional redundant underlying data, thereby improving the speed and effectiveness of model fitting~\cite{bengio2000neural, vaswani2017attention}. Leonard trains the model directly on the raw data, resulting in a large $ECT$ due to the inability to fit (loss hovers around 2). Section~\ref{sec:effect_of_components} also demonstrates the impact of hierarchical encoding on training speed and model performance. (4) Model Capability. Leonard uses multi-layer LSTM, while we chose a Single-layer Decoder-only Transformer. With the same number of parameters, the latter can better learn complex patterns and interactions among high-dimensional data processed by hierarchical encoding. Its powerful representational ability and efficient attention mechanism allow the model to fit with only a few epochs of training (generally within 4 epochs)~\cite{vaswani2017attention, jacot2018neural, lu2019understanding}.

\par
In terms of storage latency, Neo4j is the fastest with an average time of 21 seconds, followed by PostgreSQL with an average time of 45 seconds. Leonard performs the worst, with an average latency of 30,233 seconds, while \sn{} is 3,205 seconds. The low efficiency of existing model training and inference frameworks causes high time costs. Adopting advanced technology and using powerful machines can decrease costs. According to statistics, the average period of these datasets is 6.7 hours, and \sn{}'s storage latency accounts for 13.29\% of that period. Considering cold storage scenarios, the storage latency of \sn{} is acceptable. 
\par
Furthermore, we observe that the storage overhead and time costs of \sn{} don't increase linearly with the size of the original provenance graph. For example, the size of G7's provenance graph ($\sim$3470M) is 2.98 $\times$ of G5 ($\sim$1152M), but G7's $ET_{hi}$, $MMT$ ,$ECT$, and $T_{s}$ are 49.3M, 9.8M, 131M and 2,293s respectively, which are only 38\% ($\sim$128.6M), 30\% ($\sim$33.1M), 38\% ($\sim$345M), and 53\% ($\sim$4335s) of G5's. We provide a detailed analysis of this phenomenon in Section~\ref{sec:applicable_scenario}. Finally, we also observe that model capacity affects the performance of \sn{}. We provide a detailed analysis of this in Section~\ref{sec:model_selction_capacity}.

\begin{table*}[]
\centering
\caption{Time Costs of Querying at Different Depths for Different Systems. }
\label{tab:query_performance}
\scalebox{0.99}{
\begin{tabular}{c|cccc|cccc|cccc|cccc}
\hline
\textbf{Depth}                                                    & \multicolumn{4}{c|}{\textbf{D1}}                                                                                                                                                                               & \multicolumn{4}{c|}{\textbf{D2}}                                                                                                                                                                                      & \multicolumn{4}{c|}{\textbf{D3}}                                                                                                                                                                                            & \multicolumn{4}{c}{\textbf{D4}}                                                                                                                                                                                                 \\ \hline
\textbf{Quartile}                                                 & \textbf{Q1}                                       & \textbf{Q2}                                       & \textbf{Q3}                                       & \textbf{Q4}                                        & \textbf{Q1}                                       & \textbf{Q2}                                       & \textbf{Q3}                                         & \textbf{Q4}                                             & \textbf{Q1}                                       & \textbf{Q2}                                        & \textbf{Q3}                                           & \textbf{Q4}                                                & \textbf{Q1}                                       & \textbf{Q2}                                        & \textbf{Q3}                                             & \textbf{Q4}                                                  \\ \hline
\textbf{\begin{tabular}[c]{@{}c@{}}Min\\ Max\\ Mean\end{tabular}} & \begin{tabular}[c]{@{}c@{}}0\\ 0\\ 0\end{tabular} & \begin{tabular}[c]{@{}c@{}}1\\ 1\\ 1\end{tabular} & \begin{tabular}[c]{@{}c@{}}2\\ 3\\ 2\end{tabular} & \begin{tabular}[c]{@{}c@{}}3\\ 11\\ 7\end{tabular} & \begin{tabular}[c]{@{}c@{}}0\\ 0\\ 0\end{tabular} & \begin{tabular}[c]{@{}c@{}}1\\ 7\\ 4\end{tabular} & \begin{tabular}[c]{@{}c@{}}7\\ 14\\ 11\end{tabular} & \begin{tabular}[c]{@{}c@{}}15\\ 1571\\ 157\end{tabular} & \begin{tabular}[c]{@{}c@{}}0\\ 0\\ 0\end{tabular} & \begin{tabular}[c]{@{}c@{}}1\\ 13\\ 5\end{tabular} & \begin{tabular}[c]{@{}c@{}}13\\ 285\\ 77\end{tabular} & \begin{tabular}[c]{@{}c@{}}285\\ 17768\\ 2127\end{tabular} & \begin{tabular}[c]{@{}c@{}}0\\ 0\\ 0\end{tabular} & \begin{tabular}[c]{@{}c@{}}1\\ 24\\ 6\end{tabular} & \begin{tabular}[c]{@{}c@{}}24\\ 1440\\ 356\end{tabular} & \begin{tabular}[c]{@{}c@{}}2300\\ 48998\\ 17588\end{tabular} \\ \hline
\textbf{PostgreSQL (s)}                                                  & 0.45                                              & 0.45                                              & 0.45                                              & 0.44                                               & 0.42                                              & 1.65                                              & 1.72                                                & 1.62                                                    & 0.43                                              & 1.82                                               & 2.98                                                  & 3.59                                                       & 0.44                                              & 1.97                                               & 4.55                                                    & 486.60                                                       \\
\textbf{Neo4j (s)}                                                    & 0.11                                              & 0.11                                              & 0.11                                              & 0.11                                               & 0.10                                              & 0.42                                              & 0.42                                                & 0.65                                                    & 0.10                                              & 0.58                                               & 1.97                                                  & 2.39                                                       & 0.10                                              & 0.59                                               & 3.78                                                    & 67.22                                                        \\
\textbf{Dehydrator (s)}                                               & 0.01                                              & 0.31                                              & 0.30                                              & 0.31                                               & 0.01                                              & 0.37                                              & 0.61                                                & 6.17                                                    & 0.01                                              & 0.22                                               & 2.13                                                  & 99.29                                                      & 0.01                                              & 3.28                                               & 4.67                                                    & 394.68                                                       \\ \hline
\end{tabular}
}
\vspace{-0.5cm}
\end{table*}

\subsection{Querying Provenance Grpahs}
\label{sec:rq2}
To evaluate the costs of querying with \sn{}, we measured the time cost of performing queries with \sn{} and compared the results with other systems. Different configurations can affect querying efficiency, and we evaluate the query cost using the smallest model (C2 in Table~\ref{tab:models_overview}) on Dataset G1, which is a common practice~\cite{fei2021seal, ding2023leonard, hossain2020combating}. First, we randomly select 100 nodes from the provenance graph as query starting nodes. Then, we perform multiple BFSs (breadth-first searches) for each node. We refer to the number of BFS executions as the \textbf{Depth}. Many existing works~\cite{hossain2017sleuth, fang2022depimpact, zengy2022shadewatcher, king2003backtracking, hassan2020rapsheet, alsaheel2021atlas}, given a POI (Point-of-Interest), first construct a backtracking provenance graph, and this construction is a process involving multiple BFSs with time conditions. Therefore, we choose to execute the BFS algorithm on different nodes at different times to evaluate the query effectiveness of \sn{}. 
% Moreover, analysts do not expect the system to return a huge graph that is not readable~\cite{alsaheel2021atlas, hassan2019nodoze, liu2018towards}. Following the previous works, we stop searching and return the results when the number of BFSs is larger than the threshold to obtain all events that are most related to the attack. 
\par
Table~\ref{tab:query_performance} shows the time costs of querying at different depths for different systems. The first row, Depth, indicates the number of BFS executions, where we choose 1-4 (i.e., D1-D4). The second row, Quartile, represents the quartiles obtained by sorting nodes in ascending order based on the number of returned edges at that depth. For example, Q4 represents the set of nodes in the top 25\% in terms of the number of returned edges. Rows 3-5, Min/Max/Mean, show the minimum, maximum, and average values of the edges returned by the nodes in the different quartiles. Rows 6-8 represent the query times at different depths for different systems. Leonard takes several minutes to load the huge $ECT$, so we don't compare querying cost with it.
\par
As shown in Table~\ref{tab:query_performance}, Neo4j demonstrates the best overall performance ($\sim$4.92s), followed by \sn{} ($\sim$32.02s), with PostgreSQL performing the worst ($\sim$32.08s). Neo4j uses adjacency lists to store nodes and edges, making access to a node's incoming edges a fast, direct operation. On the other hand, PostgreSQL needs to scan row by row, while \sn{} needs to generate characters one by one, resulting in query efficiencies far inferior to Neo4j for both systems. We also attempted the case where Depth=5, but \sn{} required over 600s of generation time for most nodes, while PostgreSQL was unable to return results at all. Therefore, we did not include this in the table. PostgreSQL requires a significant amount of memory to store intermediate results, and when the depth is large, it encounters out-of-memory errors. \sn{}, however, does not face this issue. \textit{On queries, \sn{} has similar efficiency to PostgreSQL.} Besides, the main reason for \sn{}'s high latency is the inefficiency of existing model inference frameworks. The total query time consists of model inference time and decoding time, with model inference time accounting for over 99\% of the total. Given a query head, \sn{} constructs the $Seq$ in the form of autoregressive generation, i.e., generating each character step by step. When the length of $Seq$ corresponding to a node is large (for example, thousands), the autoregressive generation can cause high latency~\cite{vaswani2017attention, radford2018improving, socher2011parsing, yang2019xlnet}. Adopting advanced technologies and using more powerful machines can reduce this time cost.

% As shown in Table~\ref{tab:query_performance}, when depth=1, \sn{} demonstrates the highest query efficiency, with an average time cost of only 0.24s, which is 53\% and 40\% of PostgreSQL and Neo4j respectively. As the number of returned edges grows, \sn{}'s query efficiency becomes the lowest. When depth=4, \sn{}'s average time cost is 243.84s, which is 197\% and 307\% of PostgreSQL and Neo4j respectively. 

\par
In fact, \sn{} utilizes hierarchical encoding to enhance model inference speed like some works~\cite{kaiser2018fast}. As discussed in Section~\ref{sec:storing_hierarchical_encoding}, hierarchical encoding is a process that increases information density, meaning that the same information is represented with fewer characters. 
% As mentioned in Section~\ref{sec:efficient_encoding_scenarios}, when the average degree $d_{avg}>3$, hierarchical encoding can effectively reduce structure-level redundancy. 
Hierarchical encoding can reduce the number of characters that need to be generated, decreasing the number of model inferences and reducing query latency. 
% We will discuss the effectiveness of hierarchical encoding for reducing query latency in detail in Section X.X.

\subsection{Effect of Components}
\label{sec:effect_of_components}
\sn{} performs field mapping encoding (FME) on the original provenance graph to obtain $ET_{en}$, then applies hierarchical encoding (HE) on $ET_{en}$ to obtain $ET_{hi}$ and $MMT$. Finally, it inputs $ET_{hi}$ for the training model (TM) and error correction (EC) to obtain $BP_{post} (ECT+DNN+MMT)$.
\par
\begin{figure}[h!]
\centering
\subfigure[Storage Overhead]{\includegraphics[width=0.49\hsize, height=0.3\hsize]{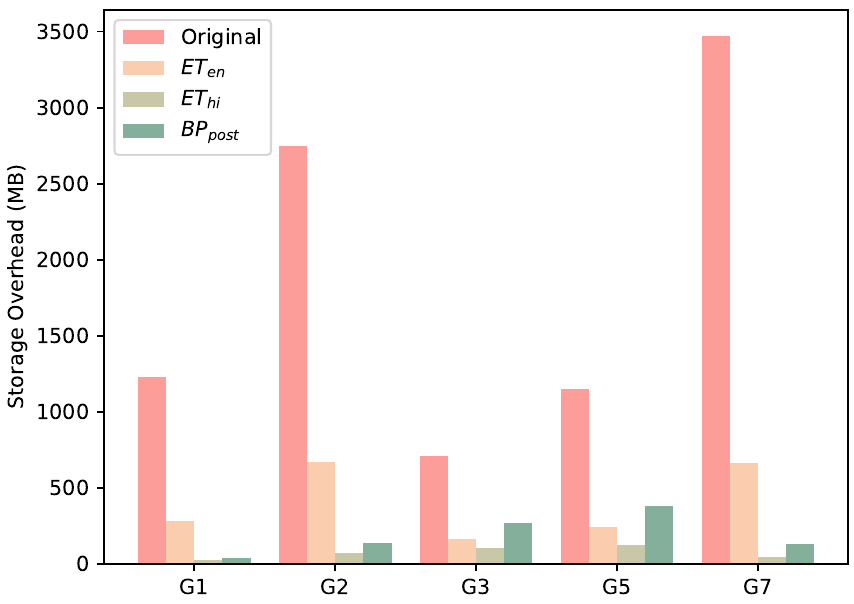}\label{fig:componets_storage}}
\subfigure[Time Costs]{\includegraphics[width=0.49\hsize, height=0.3\hsize]{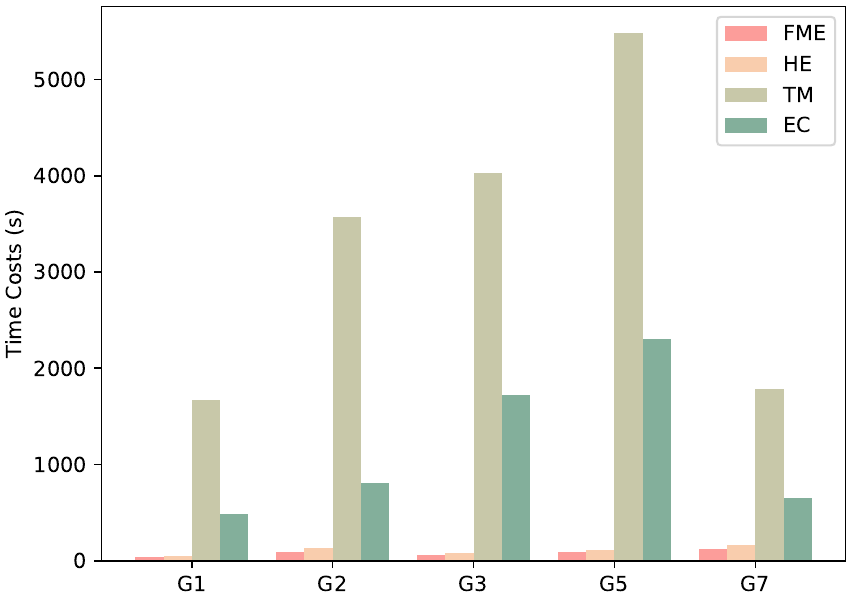}\label{fig:componets_time}}
\caption{Storage Overhead and time costs of Individual Components.}
\label{fig:componets_impact}
\vspace{-0.5cm}
\end{figure}

First, we evaluate the effect on storage overhead of three components applied in \sn{}: field mapping encoding, hierarchical encoding, and model training. Specifically, we include the original size and the sizes of storing the same provenance graph after sequentially executing field mapping encoding, hierarchical encoding, and model training. For better viewing, we choose only five datasets here, namely G1, G2, G3, G5, and G7. The results are shown in Figure~\ref{fig:componets_impact}.

\par
Figure~\ref{fig:componets_storage} shows the storage overhead for storing graphs. The average size of $ET_{en}$ is 405.2MB, which is 25.3\% of the original provenance graph ($\sim$ 1598.69MB). The average size of $ET_{hi}$ ($\sim$ 75.98MB) is 18.7\% of $ET_{en}$ and 4.7\% of the original graph. There exists a large amount of field-level and structure-level redundancy in the original provenance graph, thus both encodings achieve good compression results. After adding model training, the average storage overhead of \sn{} increases to 192.42MB, which is 12\% of the original provenance graph. Increasing the number of model parameters and training epochs could reduce the final storage overhead, but the resulting latency would be intolerable, which will be discussed in detail in Section~\ref{sec:model_selction_capacity}. Therefore, \sn{} uses model training not to further compression, but to provide a batch querying capability. Figure~\ref{fig:componets_time} shows the time costs for storing graphs. The average time costs for FME, HE, TN, and EC are 82.8s, 108.2s, 3307.8s, and 1195.6s respectively. FME and HE do not involve training, thus they are speedy. The combination of TN and EC is the model training component of \sn{}, as described in Section~\ref{sec:model_training}. 

\par
Then, we evaluate the impact of the hierarchical encoding component on model training. Specifically, we construct Dehydrator-wtHE, a version of Dehydrator without the hierarchical encoding. Intuitively, Dehydrator-wtHE directly applies the char2vec method~\cite{cao2016char2vec} to encode the strings $ET_{en}$ into numerical values $ET_{hi}$, which are then input into the model for training. Both Dehydrator and Dehydrator-wtHE use the same model (C2) on dataset G1 for the comparison experiment. Table~\ref{tab:impact_HE_onModelTrain} shows the results. Firstly, the $ET_{hi}$ size and number of segments for Dehydrator (20.19M, 21,169,175) are only 7.5\% of those for Dehydrator-wtHE (268.31M, 281,340,212). This is because HE reduces structure-level redundancy, resulting in the former's model training time (660.69s) being 8.8 $\times$ smaller than the letter (5814.42s). Furthermore, due to the low information density in Dehydrator-wtHE's $ET_{hi}$, the model is unable to fit properly, resulting in a high Loss value of 1.58. Finally, the Dehydrator's EC time (429.51s) and $ECT$ size (50.79M) are 19.48 $\times$  and 20.95 $\times$  smaller than those of Dehydrator-wtHE (8370.52s, 1064.25M), respectively. \textit{Therefore, we believe that the hierarchical encoding component can significantly improve the training speed and model performance.}

\begin{table}[]
\centering
\caption{Impact of Hierarchical Encoding on Model Training. Row 2 indicates the size of $ET_{hi}$, Rows 3 and 4 indicate the number and length of segments, Rows 5 and 6 indicate the model training time and final loss. Rows 7 and 8 indicate the error correction time and the size of $ECT$.}
\label{tab:impact_HE_onModelTrain}
\scalebox{0.99}{
\begin{tabular}{c|c|c}
\hline
\textbf{Systems} & \textbf{Dehydrator} & \textbf{Dehydrator-wtHE} \\ \hline
$ET_{hi}$ (M)       & 20.19               & 268.31                   \\
Segment Num      & 21,169,175          & 281,340,212              \\
Sgement Length   & 23                  & 8                        \\
Train (s)        & 660.69              & 5,814.42                 \\
Final Loss       & 0.88                & 1.58                     \\
EC (s)           & 429.51              & 8,370.52                \\
ECT (M)          & 50.79               & 1,064.25                 \\ \hline
\end{tabular}
}
\vspace{-0.5cm}
\end{table}
% \par
% Finally, we evaluate the impact of model selection. 

\subsection{Applicable Scenario \& Lower Bound}
\label{sec:applicable_scenario}
As described in Section~\ref{sec:effect_of_components}, field mapping encoding and hierarchical encoding are both effective in removing redundancy. However, the former applies to arbitrarily structured provenance graphs, while the latter does not. Therefore, we provide the applicable scenario for hierarchical encoding.
\par
The hierarchical encoding stage focuses on reducing structure-level redundancy, with $ET_{en}$ as input and $ET_{hi}$ as output. In a provenance graph $G = (V, E)$, the number of nodes is denoted by $n$ = $|V|$, the number of edges is denoted by $m$ = $|E|$, and the average degree is $d_{avg}$ = $\frac{m}{n}$. Furthermore, we define the space occupation in bytes for each field - $srcID$, $dstID$, $timestamp$, and $operation$ - as $BS_{s}$, $BS_{d}$, $BS_{t}$, and $BS_{o}$ respectively. Thus, the byte size of $ET_{en}$ is denoted as:  
{\setlength\abovedisplayskip{0.1cm}
\setlength\belowdisplayskip{0.1cm}
\begin{equation}
BS_{ET_{en}}=m * (BS_{s} + BS_{d} + BS_{t} + BS_{o}) 
\end{equation}}
\par
For node $v \in V$, let its number of parent nodes be $p_v$, and its number of incoming edges be $m_v$. After hierarchical encoding, the edge is structured as $e_{hi} = [v, U_{v}, startTime, endTime, MergedEdge_{v}]$, where $MergedEdge_{v} = [Operation[timeOffset[nodeOffset]]]$. Among these, the byte sizes of $v$, $startTime$, $endTime$, and $U$ are $BS_{d}$, $BS_{t}$, $BS_{t}$, and $p_{v} * BS_{s}$, respectively. To calculate the byte size of the content $BS_{con_{v}}$, we assume the number of retained unique entries in the $Operation$ field after deduplication is $m_{vo}$, the number of retained unique entries in the $timeOffset$ field under each Operation after deduplication is $m_{vot}$, and the number of retained unique entries in the $nodeOffset$ field under each $timeOffset$ after deduplication is $m_{votn}$. Then the byte size of $MergedEdge_{v}$ is:
{\setlength\abovedisplayskip{0.1cm}
\setlength\belowdisplayskip{0.1cm}
\begin{equation}
BS_{con_{v}} = BS_{o} * m_{vo}  +  BS_{t}*  \sum_{O_{v}}^{o} *m_{vot} + BS_{s}*\sum_{O_{v}}^{p}\sum_{T_v}^{t}*m_{votn} 
\end{equation}}
\par
where $\sum_{O_v}^{o}\sum_{T_v}^{t}*m_{votn} = m_{v}$. Therefore, the byte size of $ET_{hi}$ is:
% {\setlength\abovedisplayskip{0.1cm}
% \setlength\belowdisplayskip{0.1cm}
% \begin{equation}
% \begin{align}
% BS_{ET_{hi}} &= \sum_{V}^{v}(BS_{o} * m_{vo}  +  BS_{t}*(\suzm_{O_v}^{o} *m_{vot}+2)  \\
% &\quad + BS_{s}*(p_{v}+m_{v})+BS_{d})
% \end{align}
% \end{equation}
% } 
{\setlength\abovedisplayskip{0.1cm}
\setlength\belowdisplayskip{0.1cm}
\begin{align}
BS_{ET_{hi}} &= \sum_{V}^{v}(BS_{o} * m_{vo}  +  BS_{t}*(\sum_{O_v}^{o} *m_{vot}+2)  \notag \\
&\quad + BS_{s}*(p_{v}+m_{v})+BS_{d})
\end{align}
}
In Section~\ref{sec:sysdesign_preprocessing}, we performed index mapping for different fields, so these fields are all fixed and equal in size (int4 type, 32 bits). Therefore, we can abbreviate both as:
{\setlength\abovedisplayskip{0.1cm}
\setlength\belowdisplayskip{0.1cm}
\begin{equation}
\begin{aligned}
BS_{ET_{en}} &= 4*m  \\
BS_{ET_{hi}} &= \sum_{V}^{v}m_{vo} + \sum_{V}^{v}\sum_{O_{v}}^{o}m_{vot} + \sum_{V}^{v}p_{v} +3n + m
\end{aligned}
\end{equation}}
Considering that $\sum_{V}^{v}m_{vo} \le 3n$ (a node can have at most 3 operations of incoming edges, for example, a process node can have operations of Read, Sendto, and Fork), $\sum_{V}^{v}\sum_{O_{v}}^{o}m_{vot} \le m$ and $\sum_{V}^{v}p_{v} \le m$, therefore:
{\setlength\abovedisplayskip{0.1cm}
\setlength\belowdisplayskip{0.1cm}
\begin{equation}
BS_{ET_{en}} - BS_{ET_{hi}} \ge  0,\  If\ \frac{m}{n}  \ge  3
\end{equation}}
where $d_{avg}$ = $\frac{m}{n}$. Therefore, for hierarchical encoding, as long as the degree $d_{avg} \ge 3$, it constitutes an effective encoding scenario, capable of filtering structure-level redundancy and enhancing information density. Due to the high concurrency and complexity of modern operating systems, provenance graphs often have degree values far exceeding this threshold, as shown in Section~\ref{sec:eval_experimental_protocol}.
\par
We also conduct experiments to prove the above conclusion. Specifically, we generated 5 datasets in a laboratory environment, each consisting of 5 million events, with average degrees of 1, 2, 3, 4, and 5, respectively, referred to as P1-P5. We consider that $ET_{hi} < ET_{en}$ indicates that hierarchical encoding is effective. As shown in Table~\ref{tab:impact_degree_hierarchical_encoding}, hierarchical encoding is demonstrated to apply to the three datasets P3, P4, and P5. This proves that hierarchical encoding effectively filters structure-level redundancy on provenance graphs with $d_{avg} \ge 3$, as shown in our derivation above.
\par
The applicable scenario of \sn{} is for provenance graphs with $d_{avg} \ge 3$. When the $d_{avg} =1$, \sn{} exhibits a lower bound performance.

\begin{table}[]
\centering
\caption{The Impact of Degree on Hierarchical Encoding.}
\label{tab:impact_degree_hierarchical_encoding}
\scalebox{0.99}{
\begin{tabular}{c|ccccc}
\hline
\textbf{Dataset}                                                           & \textbf{P1}     & \textbf{P2}     & \textbf{P3}     & \textbf{P4}     & \textbf{P5}    \\ \hline
\begin{tabular}[c]{@{}c@{}}$BP_{pre}$ (M)\\ \# Edge\\ $ET_{en}$ (M)\end{tabular} & \multicolumn{5}{c}{\begin{tabular}[c]{@{}c@{}}304.89\\ 5,000,000\\ 96.46\end{tabular}} \\ \hline
Degree                                                                     & 1               & 2               & 3               & 4               & 5              \\
$ET_{hi}$ (M)                                                                 & 171.67          & 102.12          & 78.31           & 66.35           & 59.16          \\
Applicable                                                                 & No              & No              & Yes             & Yes             & Yes            \\ \hline
\end{tabular}
}
\vspace{-0.5cm}
\end{table}

\subsection{Model Selection and Capacity}
\label{sec:model_selction_capacity}
We consider model training as a form of utilizing the memory and batch querying capabilities of DNN models to encode provenance graphs. In this section we evaluate the impact of model selection and capacity on \sn{} respectively.
\par
\textbf{Model Selection.} As mentioned in Section~\ref{sec:model_training}, we chose the Single-layer Decoder-only Transformer (SDT) as our base model and compared it with two other models: Multi-Layer LSTM and Multi-Layer GRU. To ensure fairness, all three are compared in G1 and have similar parameter counts. Specifically, SDT has 70K parameters (C2), LSTM has 64K (Embedding Dimension=32, Hidden Dimension=64, Layer=2), and GRU has 74K (Embedding Dimension=32, Hidden Dimension=64, Layer=3). As shown in Table~\ref{tab:model_selection}, SDT's training time ($\sim660s$) is 1.29 $\times$ that of LSTM ($\sim510s$) and 1.21 $\times$ that of GRU ($\sim544s$). The generated ECT ($\sim50.79M$) is 66.50\% of LSTM ($\sim76.38M$) and 73.55\% of GRU ($\sim69.05M$). The purpose of \sn{} is to efficiently store provenance graphs. SDT has stronger feature extraction and pattern recognition capabilities, allowing it to better fit the high-density information in $ET_{hi}$, resulting in a smaller $ECT$. Therefore, we choose the Single-layer Decoder-only Transformer as the base model for \sn{}.

\begin{table}[]
\centering
\caption{Impact of Model Selection on model training and error correction.}
\label{tab:model_selection}
\scalebox{0.99}{
\begin{tabular}{c|ccc}
\hline
\textbf{\begin{tabular}[c]{@{}c@{}}Model\\ Selection\end{tabular}} & \textbf{\begin{tabular}[c]{@{}c@{}}Single-layer\\ Decode-only\\ Transformer\end{tabular}} & \textbf{\begin{tabular}[c]{@{}c@{}}Multi-layer\\ LSTM\end{tabular}} & \textbf{\begin{tabular}[c]{@{}c@{}}Multi-layer\\ GRU\end{tabular}} \\ \hline
Train (s)                                                          & 660                                                                                       & \textbf{510}                                                        & 544                                                                \\
Final Loss                                                         & \textbf{0.88}                                                                             & 1.12                                                                & 1.00                                                               \\
EC (s)                                                             & 429                                                                                       & \textbf{397}                                                        & 408                                                                \\
ECT (M)                                                            & \textbf{50.79}                                                                            & 76.38                                                               & 69.05                                                              \\ \hline
\end{tabular}
}
\vspace{-0.5cm}
\end{table}

\textbf{Model Capacity.}
Existing research has already demonstrated that the `\textit{double descent curve}' is a robust phenomenon in modern neural networks, indicating that `\textit{larger models are better}'~\cite{belkin2019reconciling, geiger2020scaling, nakkiran2021deep}. However, storage scenarios must balance latency and effectiveness. For instance, if the storage latency exceeds the period of the provenance graph, such storage becomes akin to chasing a carrot. Therefore, we designed a simple yet effective metric called \textit{Latency-to-Storage Ratio (LSR)} to evaluate the efficiency of the model training. Assuming the time cost of the storing is $T_s$, the byte size of pre-processing logs is $BS_{pre}$, and the byte size of post-processing logs is $BS_{post}$, $LSR$ is defined as:
{\setlength\abovedisplayskip{0.1cm}
\setlength\belowdisplayskip{0.1cm}
\begin{equation}
LSR = \frac{{BS_{pre} - BS_{post}}}{T} (B/s)
\end{equation}}
where $B/s$ denotes the average number of bytes compressed per second.
\par
\begin{figure}[h!]
\vspace{-0.5cm}
\centering
\subfigure[$BP_{pre}-BP_{post}$ on G1-G7]{\includegraphics[width=0.48\hsize, height=0.3\hsize]{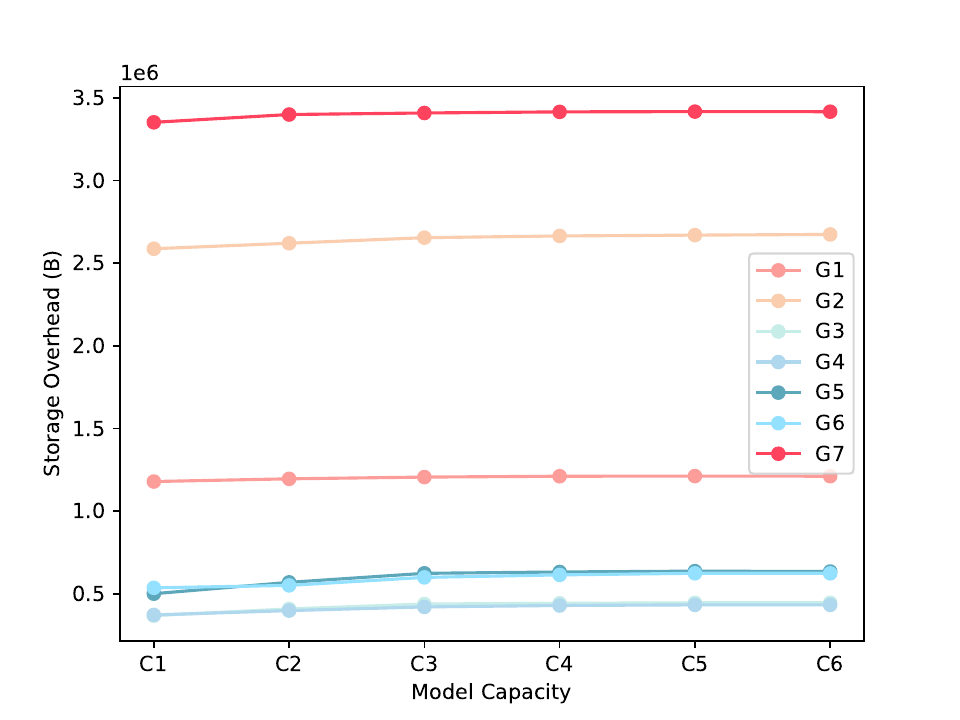}\label{fig:model_BS_pre-BS_post}}
\subfigure[$T_s$ on G1-G7]{\includegraphics[width=0.48\hsize, height=0.3\hsize]{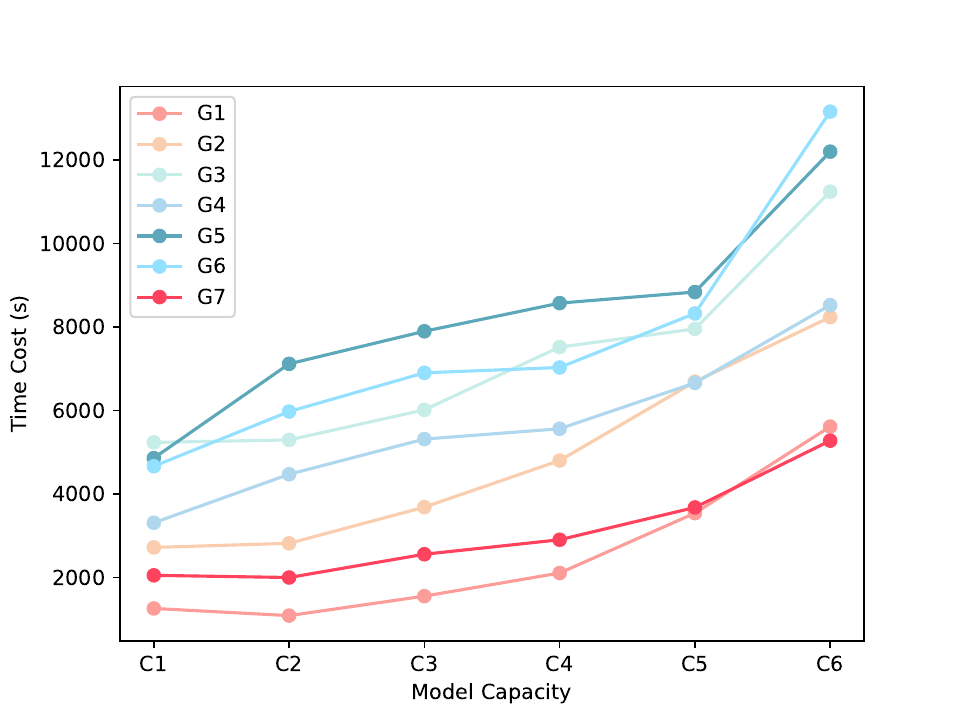}\label{fig:model_capacity_Ts}}
\subfigure[LSR (B/s) on G1, G2 and G7]{\includegraphics[width=0.48\hsize, height=0.3\hsize]{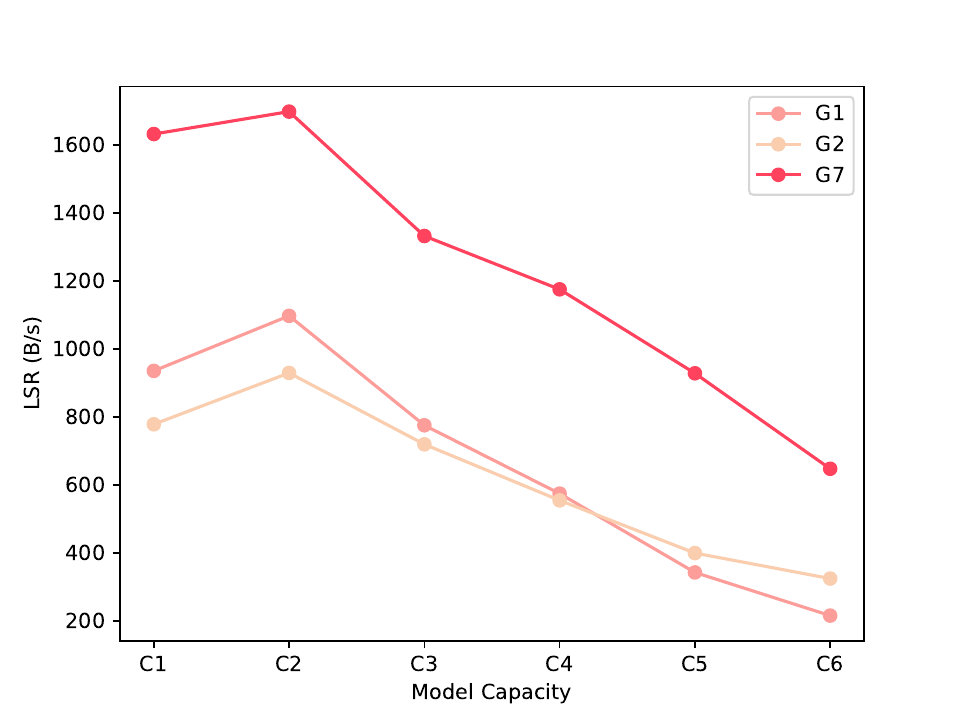}\label{fig:model_capacity_127}}
\subfigure[LSR (B/s) on G3, G4, G5 and G6]{\includegraphics[width=0.48\hsize, height=0.3\hsize]{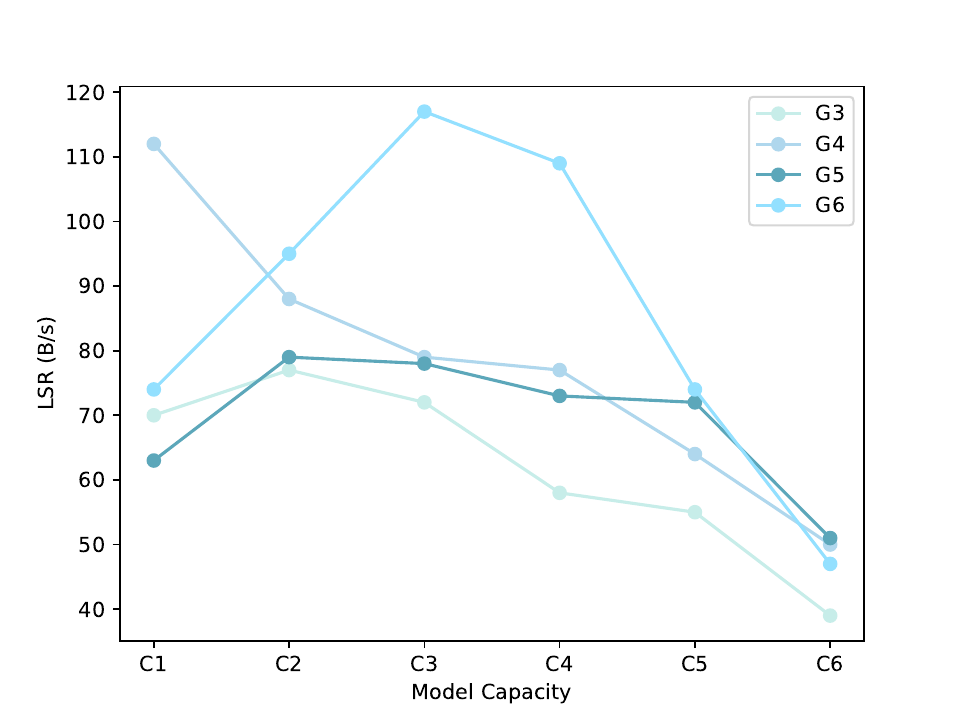}\label{fig:model_capacity_34}}
\caption{Impact of Model Capacity.}
\label{fig:model_capacity}
\vspace{-0.8cm}
\end{figure}

Given a provenance graph $G$, we observe that the model capacity $MC$ is a key variable that affects both storage overhead and latency. We construct 6 different capacities of models, as shown in Table~\ref{tab:models_overview}, and conduct evaluation experiments on datasets, recording $BS_{pre} - BS_{post}$, $T_s$, and $LSR$. Figure~\ref{fig:model_capacity} shows the result. As shown in Figure~\ref{fig:model_BS_pre-BS_post} and Figure~\ref{fig:model_capacity_Ts}, when the $MC$ increases, both storage overhead $BS_{pre} - BS_{post}$ and latency $T_s$ rise, with the latter increasing at a faster rate. 
% We therefore define $LSR$ as a function of $MC$, i.e., $LSR = Fun(MC)$. 
As depicted in Figure~\ref{fig:model_capacity_127} and Figure~\ref{fig:model_capacity_34}, the curve in the graph resembles a skewed distribution, i.e., there exists a value $\eta$ such that when $MC = \eta$, LSR reaches its maximum. For G4, $\eta=C1$, for G6, $\eta=C3$, for other provenance graphs, $\eta=C2$. \textit{Larger model is not always better, and different provenance graphs have different $\eta$.} Like most storage systems, \sn{} also needs to balance storage overhead and latency. If higher storage efficiency is prioritized, a model with larger capacity is used. If lower latency is prioritized, a model with smaller capacity is used.
% However, due to the black-box nature of DNN and the unknown distribution of provenance graphs, we can't theoretically derive the value of $\eta$. 
% However, due to the black-box nature of DNN and the unknown distribution of provenance graphs, we can't theoretically derive the value of $\eta$. Therefore we construct a \textbf{Capacity Assessment Mechanism (CAM)} based on data-driven to find the $MS$ value that approximates $\eta$.

% We observe that the ratio $R$ between the model scale ($MS$) and the provenance graph scale ($PS$) is a key variable affecting both storage latency and efficiency simultaneously. As shown in Figure X, as $R$ increases, i.e., when the model scale exceeds the provenance graph scale, both storage latency ($T$) and storage efficiency ($BS_{pre} - BS_{post}$) increase. Conversely, as $R$ decreases, both storage latency ($T$) and storage efficiency ($C_{a} - C_{b}$) decrease. Given a specific provenance graph (i.e., when $PS$ is fixed), an optimal ratio $\gamma$ exists that maximizes $LSR$. As shown in Figure X, the $LSR$ as a function of $R$ reaches a global optimum when $R = \gamma$. These observations are supported by experimental results presented in Section X. Besides, segmented storage of provenance graphs is common. For example, processing raw logs and storing provenance graphs every 24 hours. Based on the above observations, \sn{} proposes a \textbf{Capacity Assessment Mechanism (CAM)}, which pre-sets model parameters according to the provenance graph size to balance latency and efficiency, achieving the highest $LSR$. 
% \par

% \par
% First, we extract four features ($Fs$) from the provenance graph - number of edges, number of nodes, standard deviation of degree, and kurtosis of degree - as the basis for model capacity assessment. The rationale for this selection is intuitive: larger node and edge counts indicate a more complex provenance graph, necessitating greater model capacity. The standard deviation of degree and kurtosis influence the efficiency of hierarchical encoding in aggregating edge information.
% Next, we train models with varying parameter counts on this provenance graph and calculate the $LSR$.
% Finally, we repeat the above two steps for different provenance graphs and train a decision tree to fit various $Fs$ to the optimal model capacity $MS$ under the highest $LSR$. The numerical selections, effect demonstrations, and result analyses are described in detail in Section X.X. In the application phase, given a provenance graph $G$, \sn{} extracts the features $Fs$, inputs them into the decision tree, obtains the corresponding model capacity settings $MS$, and initializes the model for subsequent storage model training.

\section{Discussion}
\label{sec:6_discussion}
The purpose of \sn{} is to store provenance graphs efficiently. 
% Field mapping encoding and hierarchical encoding are both for filtering redundancy, while the DNN is for supporting batch querying. 
\sn{} saves $MMT$, $DNN$, and $ECT$ to enable event querying on the provenance graph, with $ECT$ accounting for over 90\% of the space. 
Naturally, we try to minimize $ECT$ as much as possible to further reduce storage overhead. 
$ECT$'s role is to correct the model's mispredictions, in other words, the higher the model's accuracy, the smaller the $ECT$.
Therefore, we have transformed the problem of reducing storage overhead into how to improve the model's accuracy.
\par
Intuitively, increasing the model capacity infinitely can achieve 100\% accuracy. However, like most systems~\cite{stonebraker2018cstore, lamb2012vertica}, \sn{} also faces the contradiction between performance and latency, and needs to complete the storage within an acceptable time. As shown in Section~\ref{sec:model_selction_capacity}, we prove that the larger model is not always better. For a given provenance graph, there exists an $\eta$, such that when the model capacity is $\eta$, \sn{} achieves the maximum Latency-to-Storage Ratio. Although we have evaluated models of different capacities on different provenance graphs and proposed some strategies. However, due to the black-box nature of DNN, we are unable to theoretically derive the $\eta$ for a given provenance graph. We will leave this work for the future.
\par
% Due to the slow model inference, \sn{} spends significant time on model training and autoregressive sequence generation during storage and querying, which DNN acceleration techniques can improve. 
\sn{} requires more time to store data than traditional database systems. However, this is acceptable since \sn{} is designed to store append-only provenance graphs, typically cold data. Due to its high overhead, it's important to note that \sn{} is not intended to be a general-purpose database. However, with advancements in AI acceleration techniques, \sn{} has the potential to evolve into a general graph database.
\par
Unlike traditional databases, \sn{} does not ensure ACID (atomicity, consistency, isolation, durability) properties for a sequence of operations. \sn{} is designed for tasks that involve writing data once and reading it multiple times (cold storage), i.e., the provenance graph must remain unchanged to meet integrity requirements. Since there are no multiple writers, transaction support is unnecessary. We always validate the correctness and completeness of operations when writing to the database, ensuring that the returned results are accurate as long as execution integrity is upheld. \sn{} can also be adapted for other similar use cases.

\section{Related Work}
\label{sec:7_relatedwork}
\par
\textbf{Intrusion Detection.} 
Existing provenance-based intrusion detection systems (PIDSes) can be categorized as heuristic-based and anomaly-based. Heuristic-based PIDSes~\cite{milajerdi2019holmes, milajerdi2019poirot, hossain2017sleuth, zhu2023aptshield} use empirical knowledge based on MITRE ATT\&CK's Tactics, Techniques, and Procedures (TTPs)~\cite{mitre_attck} to construct matching rules to find known attacks in the provenance graph. HOLMES~\cite{milajerdi2019holmes} develops graph-matching rules based on TTPs related to advanced persistent threat (APT) behavior. SLEUTH~\cite{hossain2017sleuth} designs a tag-based information flow propagation system that only alerts when certain confidentiality or integrity conditions are satisfied. Poirot~\cite{milajerdi2019poirot} demonstrates that these event patterns can also be mapped from the graph extracted from Cyber Threat Intelligence (CTI) reports. However, they have difficulties in detecting unknown attacks and extending scale. Anomaly-based PIDSes detect intrusions by identifying deviations from normal behavior~\cite{zengy2022shadewatcher, threatrace, cheng2023kairos, han2020unicorn, threatrace, yang2023prographer,streamspot}. Streamspot~\cite{streamspot} demonstrates a viable cluster-based modeling approach. Unicorn~\cite{han2020unicorn} visits each node to create a label to produce a histogram description of the graph which is then hashed into a fixed-length vector. The ShadeWatcher~\cite{zengy2022shadewatcher} and ThreaTrace~\cite{threatrace} take a step in identifying anomalies at the node level but fail to reconstruct complete and coherent attack stories. KAIROS~\cite{cheng2023kairos} leverages a novel graph neural network to learn the temporal evolution of the provenance graph's structural changes to quantify the degree of anomalousness for each event.

\par
\textbf{Attack Investigation.} The attack investigation is to determine the source and scope of the attack, ascertain the extent of disruption, and develop remediation and prevention strategies. RapSheet~\cite{hassan2020rapsheet} proposes a threat scoring scheme that evaluates the severity of each alert based on tactical provenance graphs (TPGs) to enable effective investigation of alerts. HERCULE~\cite{pei2016hercule} system correlates multi-source heterogeneous logs to construct a multi-dimensional weighted graph and uses the unsupervised community detection algorithm to discover attack-related paths from it. NODOZE~\cite{hassan2019nodoze} and PriorTracker~\cite{liu2018priortracker} perform statistics on historical data and assign anomaly scores to events in the dependency graph. DEPIMPACT~\cite{fang2022depimpact} calculates dependency weights globally based on multiple features and then aggregates the weights to nodes to determine suspicious points of intrusion. ATLAS~\cite{alsaheel2021atlas} uses the combination of causal analysis, natural language processing, and machine learning to establish critical patterns of attack and non-attack behavior in the dependency graph. DEPCOMM~\cite{xu2022depcomm} extracts summaries from each subgraph, enabling the generation of summary graphs from dependency graphs, thereby reducing the difficulty of investigation for analysts. WATSON~\cite{zeng2021watson} summarises the behavior of each node to infer the semantics of each audit event based on its context. 
\par
\textbf{Graph Compression.} 
Previous research has focused on developing methods to minimize log size while preserving critical forensic evidence. LogGC~\cite{lee2013loggc} employs graph analysis to identify and eliminate temporary file I/O and other "dead-end" activities, operating on the premise that events that do not contribute to the current state of the system are not useful for investigations. CPR~\cite{xu2016cpr} reduces redundant events between source and target nodes by examining \textit{interleaved flows}, meaning whether any new inputs were received at the source between the two system calls. NodeMerge~\cite{tang2018nodemerge} is based on the observation that applications often load numerous globally read-only files at launch (e.g., shared object libraries), which can be consolidated. DPR~\cite{hossain2018dpr} retains only the essential events required for accurately traversing each entity's ancestors (S-DPR) or both ancestors and successors (F-DPR), leading to a condensed provenance graph. \textit{However, these techniques are lossy.} Xie et al.~\cite{xie2011compressing, xie2012hybrid, xie2013evaluation} suggest adaptations of web graph compression and dictionary encoding techniques for provenance graphs. Their approach involves applying web graph compression methods specifically to the context of provenance graphs. Fei et al.~\cite{fei2021seal} implement the concept of Query-Friendly Compression from data mining in the realm of provenance graphs. ELISE~\cite{ding2021elise} and Leonard~\cite{ding2023leonard} integrate high-frequency field mapping encoding with deep neural networks to store 
provenance graphs efficiently.

% Besides, offline PIDSes~\cite{zengy2022shadewatcher, wang2020provdetector, han2021sigl}, due to their high operational overhead, fail to achieve efficient real-time intrusion detection. Online PIDSes~\cite{han2020unicorn, threatrace, yang2023prographer} rely on a fixed-size training set and cannot effectively address the problem of concept drift. UNICORN~\cite{han2020unicorn} combats concept drift by gradual forgetting but causes key information to decay with duration. KAIROS~\cite{cheng2023kairos} relies on the intervention of analysts to re-train the model to combat concept drift, resulting in breaking attack-agnostic and high training costs. Although KAIROS acknowledges this issue in the paper, it does not provide a targeted solution, leaving it for future work.

% \subsection{Capacitiy Assessment}
% In fact, there have been related studies exploring the relationship between model capacity and model performance.
\section{Conclusion}
\label{sec:8_conclusion}
In this paper, we propose a provenance graph storage system, \sn{}, which uses field-level mapping encoding and hierarchical encoding to filter field-level and structure-level redundancy, and utilizes DNN to support low-memory batch querying. Compared to existing databases, \sn{} achieves more efficient storage. Experimental results show that \sn{} reduces the storage space by 84.55\%. In term of storage overhead, \sn{} is 7.36 $\times$ more efficient than PostgreSQL, 7.16 $\times$ than Neo4j, and 16.17 $\times$ than Leonard.

\bibliographystyle{IEEEtran}
\bibliography{ref}

\input{appendix}

% Biography text here without a photo.

% \begin{IEEEbiography}[{\includegraphics[width=1in,height=1.25in,clip,keepaspectratio]{fig1.png}}]{IEEE Publications Technology Team}
% In this paragraph you can place your educational, professional background and research and other interests.\end{IEEEbiography}

\end{document}